\documentclass[preprint2]{aastex}

\usepackage{txfonts,amssymb,subfigure,epsfig}
\usepackage[draft]{hyperref}
\usepackage{multirow}
\usepackage{color}
\usepackage[table]{xcolor}
\usepackage{placeins}

\usepackage{natbib}
\usepackage{natbib}
\newcommand{\gsim}{\lower.7ex\hbox{$\;\stackrel{\textstyle>}{\sim}\;$}}
\newcommand{\lsim}{\lower.7ex\hbox{$\;\stackrel{\textstyle<}{\sim}\;$}}

\newcommand{\PX}{Planet Nine\,}

\newcommand{\BBa}{BB16a}
\newcommand{\BBB}{BB16b}
\newcommand{\F}{F16}

\slugcomment{Submitted to ApJ XXXX}

\shorttitle{Constraining \PX}
\shortauthors{Holman et al. 2016}

\begin{document}

\title{Observational Constraints on \PX: Cassini Range Observations}

\author{Matthew~J.~Holman and Matthew~J.~Payne}
\affil{Harvard-Smithsonian Center for Astrophysics, 60 Garden St., MS 51, Cambridge, MA 02138, USA}
\email{mholman@cfa.harvard.edu, mpayne@cfa.harvard.edu, matthewjohnpayne@gmail.com}


\begin{abstract}
We significantly constrain the sky position, distance, and mass of a possible additional, distant planet in the solar system by examining its influence on the distance between Earth and the Cassini Spacecraft.
Our preferred region is approximately centered on (RA, Dec) $=\,(40\arcdeg, -15\arcdeg)$, extending $\sim20\arcdeg$ in all directions.
\end{abstract}


\section{Introduction}\label{SECN:INTRO}
The possibility that our solar system might harbor additional, undiscovered planets has inspired generations of astronomers, from the discovery of Uranus by Herschel in (1781), to the accurate prediction of Neptune's location and its subsequent identification~\citep{Adams.1846,LeVerrier.1846a,LeVerrier.1846b}, to the discovery of Pluto by Tombaugh (1930), to the recognition that the fringes of our solar system are populated by numerous, varied populations of solar system bodies~\citep{Oort.1950,Jewitt.1993,Gladman.2001,Brown.2004}.

The outer solar system still harbors surprises and mysteries.  For example, the origins and orbits of several ``deteached'' or ``extreme scattered disk objects'' (ESDOs), such as 2001~CR105 \citep{Gladman.2002}, Sedna \citep{Brown.2004} and 2012~VP113 \citep{Trujillo.2014} cannot be explained by the solar system in its current configuration.  A number of scenarios for the formation of ESDOs, including additional planets in the outer solar system, have been proposed.  
 \citet{Trujillo.2014} noted a peculiar clustering of the ESDOs: they all have arguments of perihelion, $\omega$, near zero. 
 This does not result from known observational biases~\citep{Fuente_Marcos.2014,Gomes.2015}.

\citet{Batygin.2016} (\BBa) extended this by noting that the most distant of the ESDOs are both apsidally and {\it nodally} aligned.  That is, not only are the long axes of their orbits roughly aligned, they share the same orbital plane.  \BBa~invigorated the community with a specific hypothesis: a Neptune-mass (~$10-15 M_\earth$) planet in a distant (~$a\sim 700$~AU), eccentric ($e\sim0.6$), and inclined ($i\sim30\arcdeg$) orbit is responsible for the orbital clustering seen among ESDOs.  This planet would share the orbital plane of the ESDOs, but it would be apsidally anti-aligned.  Although a range of orbits is feasible, the principal uncertainty is the planet's location within its orbit, to which the long-term dynamics of \BBa\, are not sensitive.

Almost immediately after the appearance of the \BBa~results, \citet{Fienga.2016} (hereafter \F) demonstrated that precise range measurements to the Cassini spacecraft from ground stations on Earth can rule out  many orbital phases of \PX, assuming the nominal mass and orbit reported by~\BBa. \F~fit detailed ephemerides based on the INPOP model~\citep{Fienga.2011,Fienga.2015}, with its 150~planetary parameters and $150,000$ observations.    Certain values of true anomaly, $\nu$, (i.e. the orbital phase of \PX) result in  significantly increased residuals compared to those from fits without an additional planet. These orbital phases can be excluded.    Remarkably, the fits are {\it improved} for a range of true anomaly values.  

\F\, only evaluated a single, nominal orbit and mass for \PX.  Varying the true anomaly of the planet along its orbit simultaneously changes both the position of the object on the sky, as well as the scale of its tidal perturbation (characterized by $M/r^3$).   This makes it rather difficult to interpret and generalize the constraints.  Presumably, fits to the full observational data set with the full complement of parameters are time-consuming, precluding an exhaustive exploration of different orbits for \PX.  Even if a large number of such fits could be carried out, it be may difficult to gain intuition from models with so many free parameters.

Here we develop a simple dynamical model for the tidal influence of a distant planet on the Earth-Saturn range.  By allowing an external ephemeris, such as INPOP~\citep{Fienga.2011,Fienga.2015} or the DE432 ephemeris~\citep{Folkner.2014}, to take care of the dynamical details for us, we can focus on the terms that are most relevant to constraining \PX.   Despite its simplicity, our model accurately recovers all of the details of the more elaborate fits of ~\F.  Furthermore, our model is fast enough that we can consider the full range of sky positions, masses, and heliocentric distances to \PX.  

We organize the remainder of this paper as follows.
In section~\ref{SECN:SIMPLE} we introduce our dynamical model.  
In section~\ref{SECN:DATA} we describe the Cassini spacecraft precision ranging data.  
In section~\ref{SECN:VALIDATION} we validate our model by recovering the primary results from \F.  
Next, in section~\ref{SECN:RESULTS}, we present our all-sky constraints on the location, mass, and distance of \PX.
In section~\ref{SECN:COMBINED} we combine our results from the Cassini spacecraft ranging data with the dynamical constraints of \BBa~and \BBB. 
In section~\ref{SECN:DISC} we discuss our results, and conclude in section~\ref{SECN:CONC}.


\section{Simplified Model of Tidal Perturbations}
\label{SECN:SIMPLE}
We start with Encke's method, which considers the effect of a perturbation on a reference trajectory~\citep{Danby.1992}.  
Suppose the perturbed heliocentric position of a planet (or the Cassini spacecraft), $\vec{r}$, is given by
\begin{equation}
    \vec{r} = \vec{\rho} + \Delta\vec{r},
\end{equation}
where $\vec{\rho}$ is the planet's ephemeris position and $\Delta\vec{r}$ is the cumulative result of perturbations that were not included in the ephemeris.   Considering only the heliocentric Keplerian orbit and a perturbing acceleration, $\vec{F}$, the equations of motion are
\begin{equation}
    \frac{d^2 \vec{r}}{dt^2} = -\frac{\mu}{r^3}\vec{r} + \vec{F}
\end{equation}
and
\begin{equation}
    \frac{d^2 \vec{\rho}}{dt^2} = -\frac{\mu}{\rho^3}\,\vec{\rho}, 
\end{equation}
where $\mu=G(M_\sun + M_p)$, for a planet of mass $M_p$.  The equations of motion for the difference vector $\Delta\vec{r}$ are 
\begin{eqnarray}
    \frac{d^2 \Delta \vec{r}}{dt^2} 
    &=& 
    \mu\left(\frac{\vec{\rho}}{\rho^3} - \frac{\vec{r}}{r^3}\right) + \vec{F} \nonumber 
    \\
    &=&  
    h\,fq\,\vec{r} + h\,\Delta\vec{r} + \vec{F},
    \label{EQN:EOM}
\end{eqnarray}
where $h=\mu/\rho^3$ and
\begin{equation}
    fq = 1 - \frac{\rho^3}{r^3} \approx 3\hat{\rho}\cdot\Delta\vec{r}/\rho.
\end{equation}
The other terms, due to the interactions between planets, are all smaller by a factor $O(M_p/M_\sun)$ and will be ignored here.
Given an ephemeris or other means of obtaining $\vec{\rho}$, the perturbation $\vec{F}$, and the initial conditions $\Delta\vec{r}_0$ and $\Delta\dot{\vec{r}_0}$, $\Delta\vec{r}$, and thus $\vec{r}$, can be determined as a function of time by numerically integrating the equations of motion.

We model $\vec{F}$, the perturbation on Saturn's orbit by \PX, with a tidal potential~\citep{Hogg.1991}:
\begin{equation}
    \Phi_x = \frac{GM_X}{2 r_X^3}\left[r^2 - 3\left(\vec{r}\cdot\hat{r}_X\right)^2\right],
\end{equation}
where $\vec{r}$ is again the heliocentric position of Saturn (or Cassini) and $r=|\,\vec{r}\,|$. The mass, heliocentric distance, and heliocentric unit vector to \PX are given by $M_X$, $r_X$, and $\hat{r}_X$, respectively.  The corresponding acceleration is
\begin{equation}
    \vec{F} = \frac{GM_X}{r_X^3}\left[\vec{r} - 3\left(\vec{r}\cdot\hat{r}_X\right)\hat{r}_X\right].
    \label{EQN:TIDAL}
\end{equation}
The tidal parameter $M_X/r_X^3$ and its direction $\hat{r}_X$ completely specify the perturbation.  

The orbital period of \PX would be $18,000$ years, for its nominal semi-major axis $a\sim700$~AU.  Over the current $\sim10$~year span of the Cassini data, \PX would move only $\sim0\fdg2$.
Thus, the tidal approximation is well suited to this investigation.

\section{Data: Cassini Spacecraft Precision Ranging}
\label{SECN:DATA}

The range between the Cassini spacecraft and ground stations on Earth can be determined from the communication time between the ground stations and the spacecraft.  
In particular, the range $R$ is given by
\begin{equation}
    R = c\Delta t = | \vec{r}_s (t-\Delta t) - \vec{r}_\earth(t)|,
\end{equation}
where $c$ is the speed of light and $\Delta t$ is the elapsed time between the transmission of the signal at the spacecraft location $\vec{r}_s$ at time $t-\Delta t$ and its arrival at the ground station location $\vec{r}_\earth$ at time $t$.

Figure~\ref{FIG:COMPARISON} shows the residuals $\delta R$ of the distance between Cassini and Earth, from the ephemeris fits of \F.
The residual
\begin{equation}
    \delta R_i = R_{o,i} - R_{c,i}
\end{equation}
is the difference between the observed range $R_{o,i}$ and the range calculated from the ephemeris $R_{c,i}$ for the $i$-th observation.  The residuals in figure~\ref{FIG:COMPARISON} come from the best fit of \F\, without an additional planet.

Rather than doing full ephemeris fits to get the ranges, which involves calculating $R_{c,i}$ and comparing to $R_{o,i}$, we match the {\it range residuals}, which only involves calculating the difference vectors $\Delta\vec{r}$ and accessing an ephemeris to obtain the corresponding reference vectors $\vec{\rho}$.

Here we introduce our simplifying assumptions.  As already stated, the range observations measure the light travel time from Cassini to the ground station.  However, the distance is dominated by the separation between the barycenter of the Saturn system and the geocenter.  Both the position of Cassini with respect to the Saturn system barycenter and the position of the ground stations with respect to the geocenter are well known and well modeled~\citep{Fienga.2016}, as is the mass of the Saturn system.  Furthermore, the orbit of the Earth is known precisely from VLBI observations of distant sources~\citep{Folkner.2014}.  Thus, we assume (1) that the Cassini range residuals are primarily due to the {\it unmodeled} acceleration of the Saturn system barycyenter, (2) that we can ignore the acceleration of the geocenter due to \PX, and (3) that Cassini range residuals are well approximated by the changes in the separation between the barycenter the Saturn system and the geocenter.  
We validate our assumptions in section~\ref{SECN:VALIDATION}.

We model the observations from $\vec{\rho}_i$, $\Delta\vec{r}_i$, and $\vec{r}_{\earth,i}$ (the position of the geocenter), where $i$ references the $i^{\rm th}$ observation.   
We add the acceleration $\vec{F}$ from \PX  and fit for the values of $\Delta\vec{r}_0$ and $\Delta\dot{\vec{r}}_0$ that minimize the residuals. 
This very accurately approximates the result of the full ephemeris fits of \F, as demonstrated in Figures~\ref{FIG:COMPARISON}~and~\ref{FIG:COMPARISON2}. 

The new observable is the range residual 
\begin{eqnarray}
\delta R_i =|\,\vec{\rho}_i +\Delta\vec{r}_i-\vec{r}_\earth\,|-|\,\vec{\rho}_i-\vec{r}_\earth\,| \approx\hat{R}_i \cdot\Delta\vec{r}, \nonumber   
\end{eqnarray}  
where
$\hat{R}_i = (\vec{\rho}_i-\vec{r}_\earth)/|\,\vec{\rho}_i-\vec{r}_\earth\,|$
is the unit vector from the Earth to Saturn.
We directly fit the  range residuals by minimizing
\begin{equation}
    \chi^2 = \sum_i \left(\frac{1}{\sigma_p^2}\right)\left[\delta R - \hat{R}_i \cdot\Delta\vec{r}_i\right]^2,
    \label{EQN:CHI}
\end{equation}
where $\sigma_p$ is the uncertainty of the range measurements and the sum is over all observations.
Specifically, we vary the six components of $\Delta\vec{r}_0$ and $\Delta \dot{\vec{r}}_0$ to minimize $\chi^2$.    Although fits that include the parameters of other planets might absorb slightly more of the residuals, the six components we consider evidently account for most of the leverage of the free parameters (see Figures~\ref{FIG:COMPARISON}~and~\ref{FIG:COMPARISON2}).

Our model is built from the {\it Orbfit} package of \citet{Bernstein.2000} and features the following modifications.
\begin{itemize}
\item It integrates the equations of motion given by equation~\ref{EQN:EOM} with a leap frog symplectic integrator.   (Note that the heliocentric motion has already been accounted for.)

\item It obtains the positions of the geocenter and the Saturn system barycenter from the DE432 ephemeris.

\item The six components of $\Delta\vec{r}_0$ and $\Delta\dot{\vec{r}}_0$, the parameters to be fitted, are initialized to zero.  

\item The minimization is done with the Levenberg-Marquardt algorithm~\citep{Press.1986}.  The equations of motion are time-dependent but nearly linear in $\Delta\vec{r}$.  Thus, Levenberg-Marquardt is well behaved.

\item The derivatives of each observed range residual, $\hat{R}_i \cdot\Delta\vec{r}_i$, with respect to the fitted parameters, are required by the algorithm and are determined numerically.

\end{itemize}

\section{Model Validation}
\label{SECN:VALIDATION}

    \begin{figure*}[thp]
    \begin{minipage}[b]{\textwidth}
    \centering
    \includegraphics[trim = 0mm 0mm 0mm 0mm, clip, angle=0, width=0.85\textwidth]{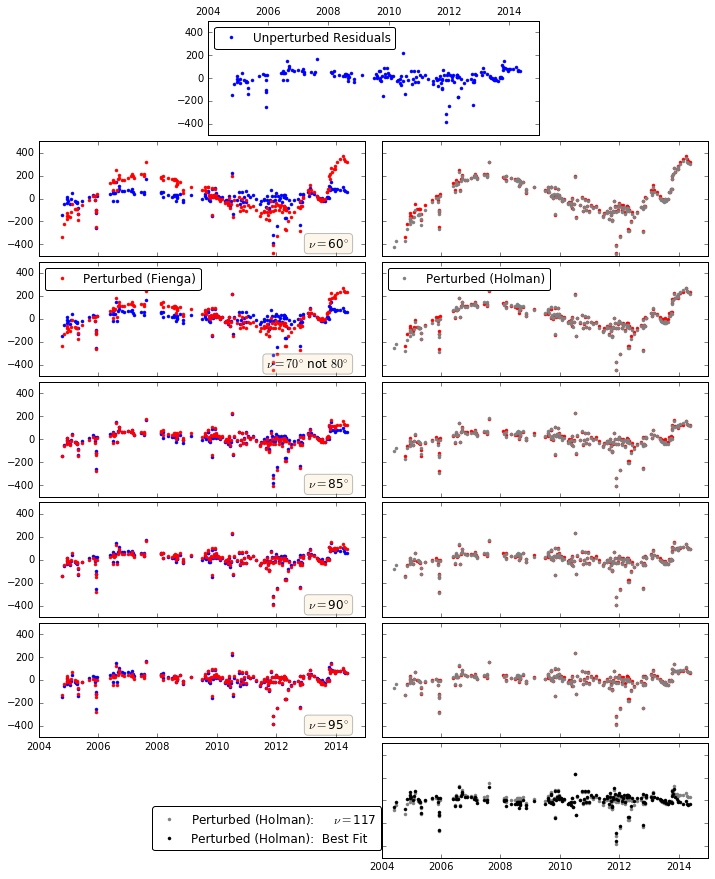}
    \caption{ %
    Comparison of the results from \F\, with those of our simplified model. 
    {\bf Top: } The unperturbed model of \F.
    {\bf Left:} we reproduce Figure 1 from \F, illustrating the differences between their perturbed models (red) and their unperturbed model (blue) for different values of the orbital true anomaly (assuming the nominal orbit for \PX as described in Section \ref{SECN:VALIDATION}).
    {\bf Right:} we compare the perturbed models from \F\, (red) with our simplified perturbation model (gray) for the same true anomalies, illustrating the striking similarity between the results.
    At the bottom-right we plot our simplified perturbation model results for the best-fit model of \F\, which lies on the nominal orbit (gray) and our overall best fit model (black) which occurs for $RA,DEC = 170\arcdeg, -57\arcdeg$ which lies away from the nominal orbit.
    Here we are plotting \emph{all} data points, \emph{including} statistical outliers. 
    We believe that panel 2 of Figure~1 in \citet{Fienga.2016} was mislabelled and that the data were actually for $\nu=70^{\circ}$ (rather than $\nu=80^{\circ}$: see Appendix \ref{APP:FIENGA} for further discussion). 
    }
    \label{FIG:COMPARISON}
    \end{minipage}
    \end{figure*}
    %

Our results rely entirely upon the residuals from the detailed, unperturbed ephemeris model of \F.  
Without their careful work, as well as their insight into the sensitivity of the Cassini range measurements to the gravitational perturbations from \PX, we would not be able to carry out the present investigation.
Furthermore, if the ephemeris models of \F\, are incorrect, our results will be too.

We re-emphasize that we are using the {\it range residuals} of \F\, rather than the range measurements themselves.  
We extract the range residuals as a function of time directly from the plots in Figure~1 of \F.  
The range residuals from their unperturbed (i.e. without \PX) model form our primary data set.  
We show this in the top panel of Figure~\ref{FIG:COMPARISON}.  
On the left hand side of Figure~\ref{FIG:COMPARISON}, we reproduce Figure 1 from \F, illustrating the differences between their perturbed models (red) and their unperturbed model (blue) for several values of the orbital true anomaly, assuming the nominal orbit for \PX as described below and illustrated in Figure \ref{FIG:SIMPLE_COMBO}.
On the right-hand side of Figure~\ref{FIG:COMPARISON}, we re-plot the range residuals for their various perturbed models (red) along with our simplified perturbation model (gray) for the same true anomalies, demonstrating the striking similarity between the results, and validating of our approach.
We believe that panel 2 of Figure~1 in \citet{Fienga.2016} was mislabelled and that the data were actually for $\nu=70^{\circ}$ rather than $\nu=80^{\circ}$ (see Appendix \ref{APP:FIENGA} for further discussion and verification).

    \begin{figure}[thp]
    \centering
    \includegraphics[trim = 0mm 0mm 0mm 0mm, clip, angle=0, width=1.0\columnwidth]{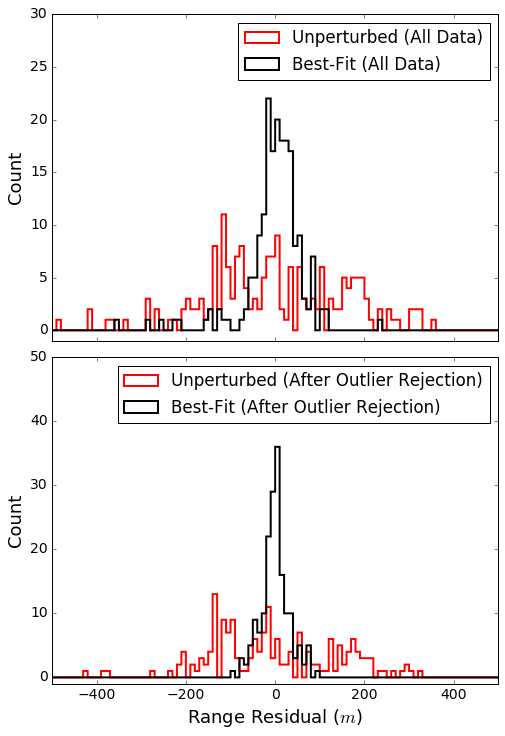}
    \caption{ %
    Outliers in the range residuals. 
    All histograms have bin-width $10\,m$.
    {\bf Top: } We plot the distribution of the basic range residuals for an unperturbed model (red line) and then show the best-fit perturbed model based on these basic range residuals (black line). It can be seen that this black line has a number of outlying values, predominantly negative, as expected from a visual examination of Figure \ref{FIG:COMPARISON}.
    {\bf Bottom: } We clip at the $3\,\sigma$ level (as described in Section \ref{SECN:VALIDATION}) to remove the outliers, resulting in the range residuals for an unperturbed model after rejection of outliers as plotted using a thick red line. When we use this as a basis for fitting, the best-fit perturbed model has a compact distribution of range residuals (thick black line), close to the desired gaussian form.
    We use the outlier-rejected residuals as the basis for all subsequent analysis. 
    }
    \label{FIG:HIST}
    \end{figure}
    %

\F\, report an RMS of $\sim75$~m for the residuals of their unperturbed fits to the Cassini range measurements.  
All panels in Figure~\ref{FIG:COMPARISON} \emph{include} statistical outlier data points.  
In the top panel of Figure~\ref{FIG:HIST} we show histograms of the residuals for both the unperturbed and global best fit models using all data points, including statistical outliers.  
The outliers are clearly evident in Figure~\ref{FIG:COMPARISON} and in the histograms of Figure~\ref{FIG:HIST}.  

We identify and eliminate the outliers as follows.
We start with the global best fit to the data with no outliers rejected, following the process described in the next section.   
In each iteration we compute the standard deviation of the residuals and reject any points that are more than $3\sigma$ from the mean.  
For this data set, the process converged in two iterations, resulting in a total of 15 points rejected and a standard deviation of $31.95$~m.   
The distribution of residuals \emph{after} the elimination of outliers is plotted in the bottom panel of Figure \ref{FIG:HIST}. 
After rejecting the outliers the distribution of residuals for the best fit model is well represented by a normal distribution.
For the remainder of our analysis, we adopt $\sigma=31.95$~m as the uncertainty of the range residuals.


    \begin{figure}[thp]
    \centering
    \includegraphics[trim = 0mm 0mm 0mm 0mm, clip, angle=0, width=\columnwidth]{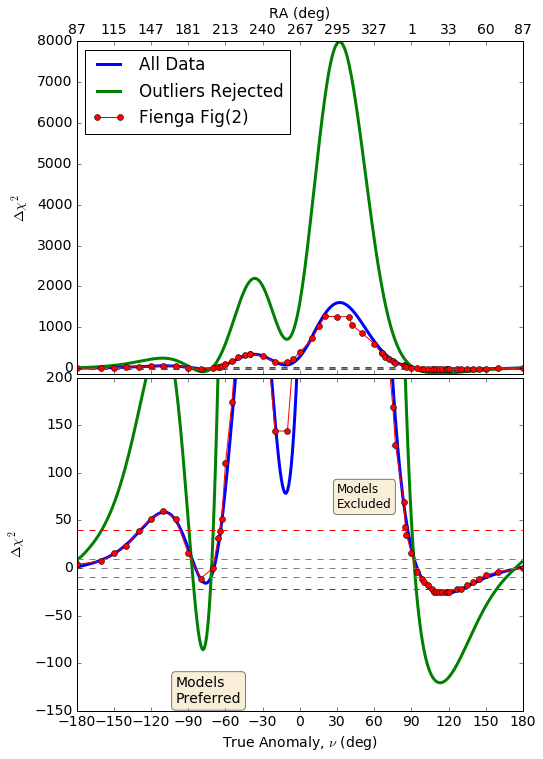}
    \caption{ %
    Comparison of the results from \F~with those of our simplified model, illustrating the $\Delta\chi^2$ as a function of true anomaly, $\nu$, for the nominal orbit analysed by \F. 
    We illustrate in blue the results obtained from performing our analysis using all Cassini range-residual data.
    In green we plot the results obtained when we reject outlying points.
    In red we plot the results from \F~(converted from $\Delta\,RMS$ to $\Delta\chi^2$).
    Gray horizontal lines are $\Delta\chi^2=\pm9$, equivalent to $3\sigma$ for a 1-degree of freedom model. 
    Red horizontal lines are the $\Delta\sigma$ criteria used by \citet{Fienga.2016} (see Appendix \ref{APP:FIENGA:CHI}).
    %
    Our results \emph{without} rejection are very similar to those of \F~(red).
    Removing outlying data points (green) allows a ``cleaner'', more decisive measurement, as the statistical significance of all regions becomes enhanced. 
    }
    \label{FIG:COMPARISON2}
    \end{figure}
    %
    
\F\, analyze the perturbations arising due to a specific single nominal orbit from \BBa, which they take as being for a planet with 
$M=10M_{\oplus}$, 
$a=700\,AU$,
$e=0.6$, 
$i=\,30\arcdeg$,
$\Omega=113\arcdeg$, and
$\omega=150\arcdeg$,
where these quantities are the mass, semi-major axis, eccentricity, inclination, longitude of ascending node, and argument of pericenter, respectively.
The true anomaly, or equivalently the mean anomaly, is unconstrained.  Thus, \F\, sample around the orbit, implicitly sampling varying distances and positions on the sky, and examine the quality of the fit at each location. 
We reproduce their analysis using our range-residual methodology, and plot a comparison of our results with those of \F\, in our Figure \ref{FIG:COMPARISON2}.

We extract the relevant data (red points) from Figure 2 in \F, which illustrate the goodness-of-fit statistic as a function of true anomaly, $\nu$.
We plot a comparison in Figure~\ref{FIG:COMPARISON2} between the results of \F\, and our own results with all data (blue) and with outlier data rejection (green).
We note that \F~use an unconventional measure to decide the statistical significance of their perturbed fits to the Cassini data, taking the normalized difference in the RMS value of the residuals for their various perturbed fits, $\sigma_p$ and comparing it to their unperturbed RMS of the residuals, $\sigma_0$, and plotting the quantity, $(\sigma_p - \sigma_0)/\sigma_0$ as a function of true anomaly.
As such we convert their measure to a $\chi^2$ measure via $\chi^2=N\sigma_p^2/\sigma_0^2$ to make it directly comparable to our own. 
We provide further discussion of the statistical measure used by \F~in Appendix \ref{APP:FIENGA}.

We draw a number of conclusions from Figure~\ref{FIG:COMPARISON2}:
\begin{itemize}
    \item We accurately reproduce the results of \F~if we use all range residual data;
    \item Removing outlying data points (green) allows a  more decisive measurement, as the statistical significance of all regions becomes enhanced, hence we use the outlier-rejected residuals as the basis for all subsequent analysis; 
    \item Even using all data points (no rejection), the metric used by \cite{Fienga.2016} errs  on the conservative side.  There are ranges of true anomaly (e.g. $-160\arcdeg<\nu<-130\arcdeg$) for which \cite{Fienga.2016} stated that the data are unable to constrain the presence of\PX, but for which we believe the data meaningfully exclude \PX (for that orbit); 
    \item We further confirm that the point labelled $\nu=80\arcdeg$ in Figure 1 of \F~is likely mislabelled and is instead $\nu=70\arcdeg$ (see Appendix~\ref{APP:FIENGA} for details.)
\end{itemize}

In summary, we find that our range-residual method accurately and rapidly reproduces the results of \F. 
Moreover, we find that there is significant room for further investigation, as excluding outlying data points and using a standard $\chi^2$ statistic allow for more nuanced model evaluation.

\section{Results: Constraints From Cassini Ranging Data}
\label{SECN:RESULTS}

In Section \ref{SECN:VALIDATION}, in which we demonstrated our model by first reproducing the primary results of~\F, we focused on a specific nominal mass and orbit for \PX. 
We now relax the assumption that \PX must have these specific masses and orbits, and instead we complete a general investigation in which a \emph{tidal} perturber (as described in Section \ref{SECN:SIMPLE} and Equation \ref{EQN:TIDAL}) is placed at varying locations on the sky. 
For each value of the tidal perturbation, $M\,r^{-3}$, and location on the sky, we minimize $\chi^2$ as defined in Equation \ref{EQN:CHI}.

We use the HEALPix tiling of the sky \citep{Gorski.2005} to test a complete set of perturbing planet locations, uniformly distributed on the sky.
For the results presented in Section \ref{SECN:RESULTS}, we concentrate on the $N_{\mathrm{side}}=2^5=32$ resolution level, resulting in 12,288 tiles.
We interpret each HEALPix tile as representing an individual possible perturber position in an equatorial frame. 
For a given HEALPix tile, i.e a given RA \& Dec, we study a range of tidal perturbations for \PX.  

The combination of two sky coordinates and one tidal parameter means that we have three degrees of freedom.  Hence, a $\chi^2$ model assuming gaussian errors would have $1,\,2,\,\&\,3\sigma$ significance for $\Delta\chi^2=3.5,\,8.0,\,\&\,14.2$ respectively.

We use the range residuals with \emph{outliers rejected} and an RMS of $\sim32\,$m as described in Section \ref{SECN:VALIDATION} and which lead to the results in the lower panel of Figure \ref{FIG:HIST}.

\subsection{Perturbed Orbital Fits}
\label{SECN:RESULTS:TIDAL}
\begin{figure}[thp]
    \centering
    \includegraphics[trim = 0mm 0mm 0mm 0mm, clip, angle=0, width=1.0\columnwidth]{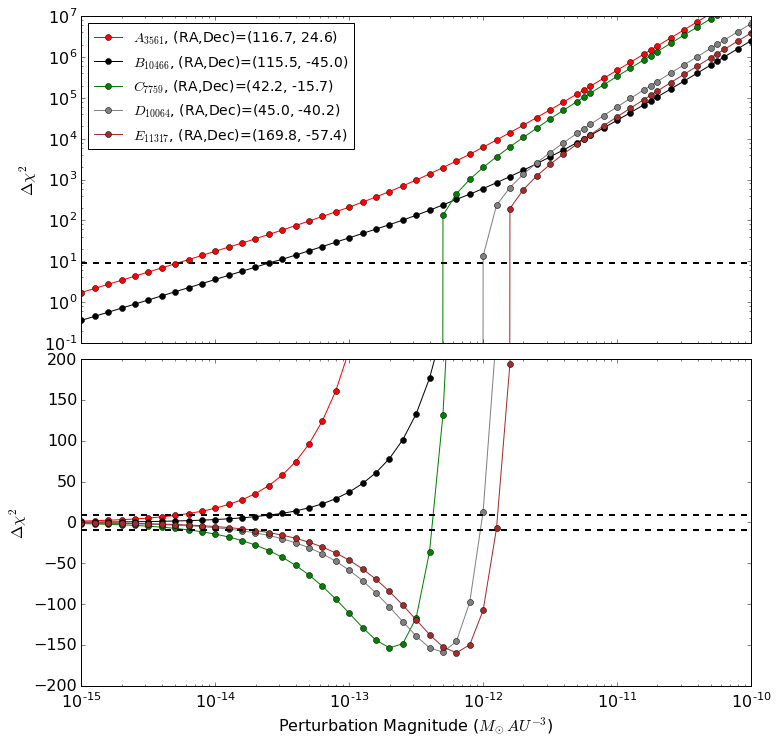}
    \caption{ %
    We plot the dependence of $\chi^2$ on perturbation magnitude, $M/r^3$ for five different healpix regions.
    The red and black lines exhibit purely positive $\Delta\chi^2$ values, rising as the perturbation magnitude increases.
    In contrast, the green, gray, and brown lines exhibit a local minimum in $\Delta\chi^2$ where the perturbations significantly improve the fits.
    With reference to Figure \ref{FIG:SKY}:
    The red line at (RA, Dec) $=$\,$(117\arcdeg, 25\arcdeg)$ is within the strip of preferred orbits from \BBB; 
    The black line at $(116\arcdeg, -45\arcdeg)$ is outside the strip of preferred orbits from \BBB; 
    The green line at $=$\,$(42\arcdeg, -16\arcdeg)$ is highly significant and located within the strip of preferred orbits from \BBB; 
    The gray line at $(45\arcdeg, -40\arcdeg)$ is in the highly significant minimum just below the strip of preferred orbits from \BBB; 
    The brown line at $(170\arcdeg, -57\arcdeg)$ is the global best-fit from our work and is far outside the strip of preferred orbits from \BBB. 
    The dashed horizontal lines indicate the $\Delta\chi^2$ change required such that the change would be inconsistent with an unperturbed model at $3\,\sigma$ confidence.
    }
    \label{FIG:LINE}
    \end{figure}

    To illustrate the dependence of our results on both the scale of the tidal perturbation and the location of the perturber on the sky, we plot in Figure~\ref{FIG:LINE} the results obtained for five particular HEALPix locations.
    
\begin{figure*}[thp]
    \begin{minipage}[b]{\textwidth}
    \centering
    \includegraphics[trim = 0mm 0mm 0mm 0mm, clip, angle=0, width=0.95\textwidth]{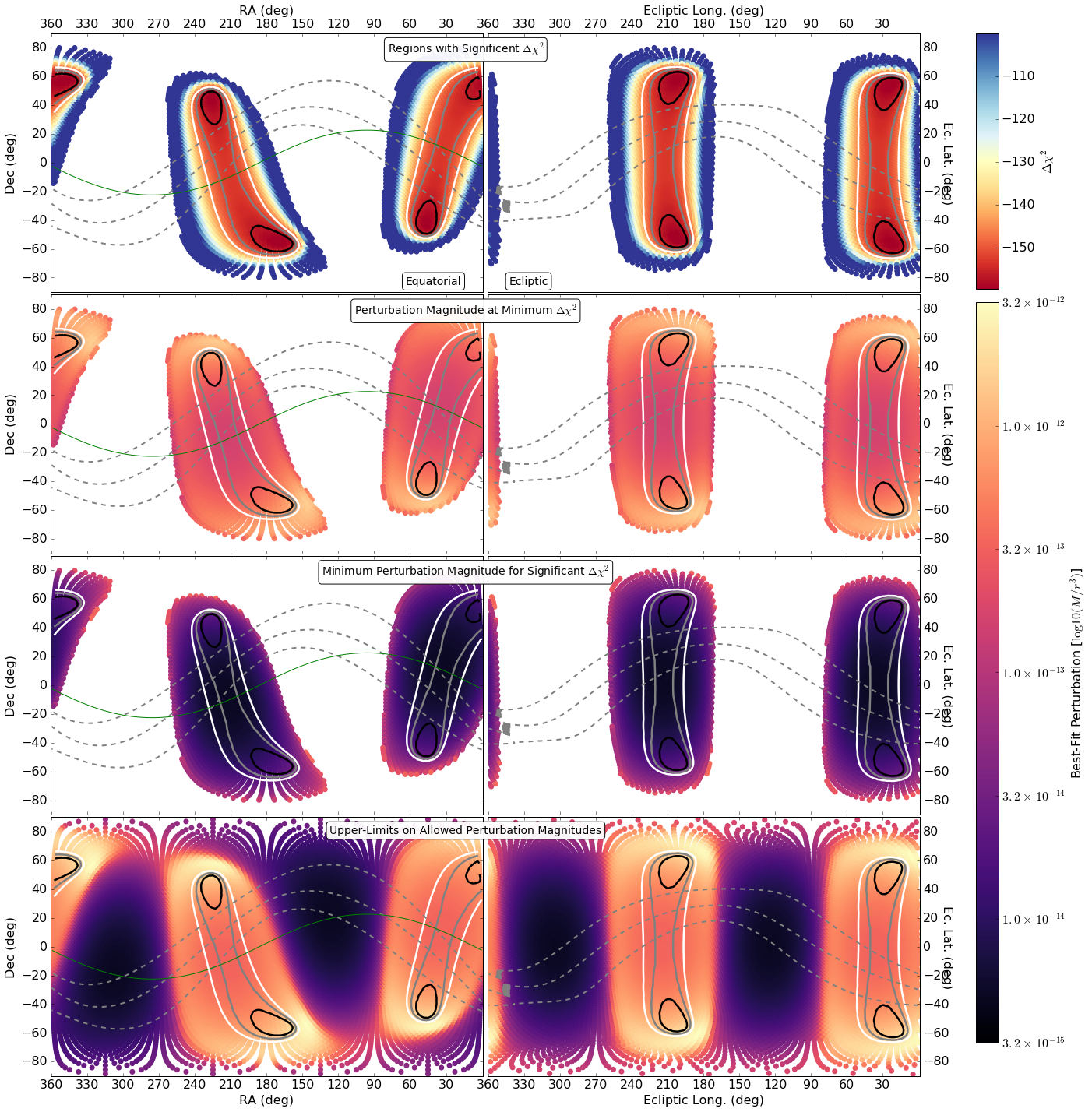}
    \caption{ %
    Constraints on the perturbation magnitude as a function of sky coordinates.
    {\bf Top:} Regions of the sky which exhibit a local minimum of the kind seen in the lower plot of Figure \ref{FIG:LINE}, illustrating the depth of the $\Delta\chi^2$ that is reached.
    {\bf 2nd-Top:} The perturbation magnitude which leads to the minimum $\Delta\chi^2$ plotted in the top plot.
    {\bf 2nd-Bottom:} The lower limit to the perturbation magnitude which still gives a significantly improved fit, i.e. the left-hand side of the minima in the lower plot of Figure \ref{FIG:LINE}.
    {\bf Bottom:} The upper limit to the perturbation magnitude, i.e. the strength of perturbation which raises the $\Delta\chi^2$ above the horizontal line in the upper plot of Figure \ref{FIG:LINE}. 
    On the {\bf left} all coordinates are equatorial, on the {\bf right} all coordinates are ecliptic.
    Black, gray and white contours are at 
    $1\sigma,\,2\sigma,\,\&\,3\sigma$ respectively. 
    To guide the eye, we repeat these contours in all panels.
    The gray, dashed lines are approximate minimum, median and maximimum declination ranges as a function of RA for the preferred range of orbits outlined in \BBB\, and explicitly plotted in our Figure \ref{FIG:SIMPLE_COMBO}.
    The green line is the orbit of Saturn.
    For half the sky, perturbations $\gsim10^{-14}\,M_{\odot}\,AU^{-3}$ can be ruled out (purple and black regions, bottom plot), while for the other half of the sky significant improvements to the fit are produced by perturbation magnitudes in the range $10^{-13}\,-\,10^{-12}\,M_{\odot}\,AU^{-3}$.
    In particular, we find that the best fits are confined (to $3\sigma$) within two thin ($\sim40\arcdeg$-wide) stripes perpendicular to Saturn's orbit. 
    }
    \label{FIG:SKY}
    \end{minipage}
    \end{figure*}

    The red line is for a HEALPix tile at (RA, Dec) $=$\,$(117\arcdeg, 25\arcdeg)$, which is within the strip of preferred orbits from \citet{Brown.2016} (\BBB) (see Figure \ref{FIG:SKY} for an all-sky plot); 
    The black line is for a tile at  $(116\arcdeg, -45\arcdeg)$, which is outside the strip of preferred orbits from \BBB; 
    The green line is for a tile at  $(42\arcdeg, -16\arcdeg)$ that is highly significant and located within the strip of preferred orbits from \BBB; 
    The gray line is for a tile at  $(45\arcdeg, -40\arcdeg)$ that is in the highly significant minimum just below the strip of preferred orbits from \BBB; 
    The brown line is for a tile at  $(170\arcdeg, -57\arcdeg)$ that is the global best-fit location from our work and is far outside the strip of preferred orbits from \BBB; 
    
    In general, locations such as those at (RA, Dec) $=$\,$(117\arcdeg, 25\arcdeg)$ and $(116\arcdeg, -45\arcdeg)$ (red and black lines, respectively) exhibit purely positive $\Delta\chi^2$ values, rising as the perturbation magnitude increases.
    A perturber placed at such a location never improves the fit, and can be ruled out to a perturbation magnitude $M\,r^{-3}\sim10^{-14}M_{\odot}\,AU^{-3}$.
    
    In contrast, regions such as those at (RA, Dec) $=$\,$(42\arcdeg, -16\arcdeg)$, $(45\arcdeg, -40\arcdeg)$ and $(170\arcdeg, -57\arcdeg)$ (green, gray and brown lines respectively) all exhibit highly significant  minima in $\Delta\chi^2$, allowing us to:
    (a) find the perturbation associated with the minimum in $\Delta\chi^2$;
    (b) find the smallest perturbation which causes a significant ($|\Delta\chi^2| < 14.2$) improvement in fit (i.e. the left-hand side of the minima curves); 
    as well as 
    (c) rule out large perturbations which only degrade the fit. 

    %

    We repeat the approach illustrated in Figure \ref{FIG:LINE} and apply it to the entire sky, plotting our results in Figure \ref{FIG:SKY}.
    
    In the top panel of Figure \ref{FIG:SKY} we use a color-map to illustrate those regions of the sky which exhibit any significant $\Delta\chi^2$ minimum (of the kind plotted in green, gray and brown in Figure \ref{FIG:LINE}).
    White corresponds to sky locations that do \emph{not} exhibit any statistically significant $\Delta\chi^2$ minimum.
    We plot contours  in black, gray and white corresponding to $1,\,2,\,\&\,3\sigma$, respectively.
    To guide the eye, we also plot the equatorial coordinates of Saturn (green) and the swath of proposed orbits for \PX from \BBB\, (further details are provide in Section \ref{SECN:COMBINED} to describe exactly what orbital parameters we assumed for this swath). 
    
    Our results are extremely constraining, with our best-fit locations being confined to very small regions of the sky. 
    Moreover, the regions of sky favored by our Cassini analysis are approximately orthogonal to the preferred orbital swath from \BBB, allowing for a significant reduction in the area on the sky where \PX could be located.
    We note that there are two equal-and-opposite regions on the  sky due to the symmetric nature of the tidal perturbation model. 
    
    The second panel of Figure \ref{FIG:SKY} plots the perturbation which minimizes  $\chi^2$ for each HEALPix tile.  This is confined to a range 
    \begin{eqnarray}
    2\times10^{-13}\lsim\,M\,r^{-3}\lsim\,1\times10^{-12}\,M_{\odot}\,AU^{-3}.\nonumber
    \end{eqnarray}
    
    Smaller perturbations (than those of panel 2) can still significantly improve $\chi^2$.  In panel three of Figure \ref{FIG:SKY} we illustrate the smallest perturbation scale that can still cause a significant improvement to the fit at a given location.  This is typically
    \begin{eqnarray}
    2\times10^{-15}<M\,r^{-3} < 3\times 10^{-14}\,M_{\odot}\,AU^{-3}.\nonumber
    \end{eqnarray}
    
    Finally, for any given location on the sky, we can rule out perturbers above a certain tidal parameter by finding when the increase in $\chi^2$ becomes too large to have occurred by chance. We plot this measure in the bottom panel of Figure \ref{FIG:SKY}.  For large areas on the sky, we can rule out any perturbation 
    \begin{eqnarray}
    M\,r^{-3} \gsim 10^{-14}\,M_{\odot}\,AU^{-3}.\nonumber
    \end{eqnarray}
    
    All of the plots in Figure \ref{FIG:SKY} are repeated, with the second set of plots (on the right-hand side) being plotted in the plane of the ecliptic (Saturn's inclination is only $2\fdg49$).  The areas on the sky that we constrain are thin stripes, perpendicular to the plane of Saturn's orbit. 

    %

\subsection{Summmary of Cassini-Only Results}
\label{SECN:CASS:SUMM}

\begin{figure}[thp]
    \centering
    \includegraphics[trim = 0mm 0mm 0mm 0mm, clip, angle=0, width=\columnwidth]{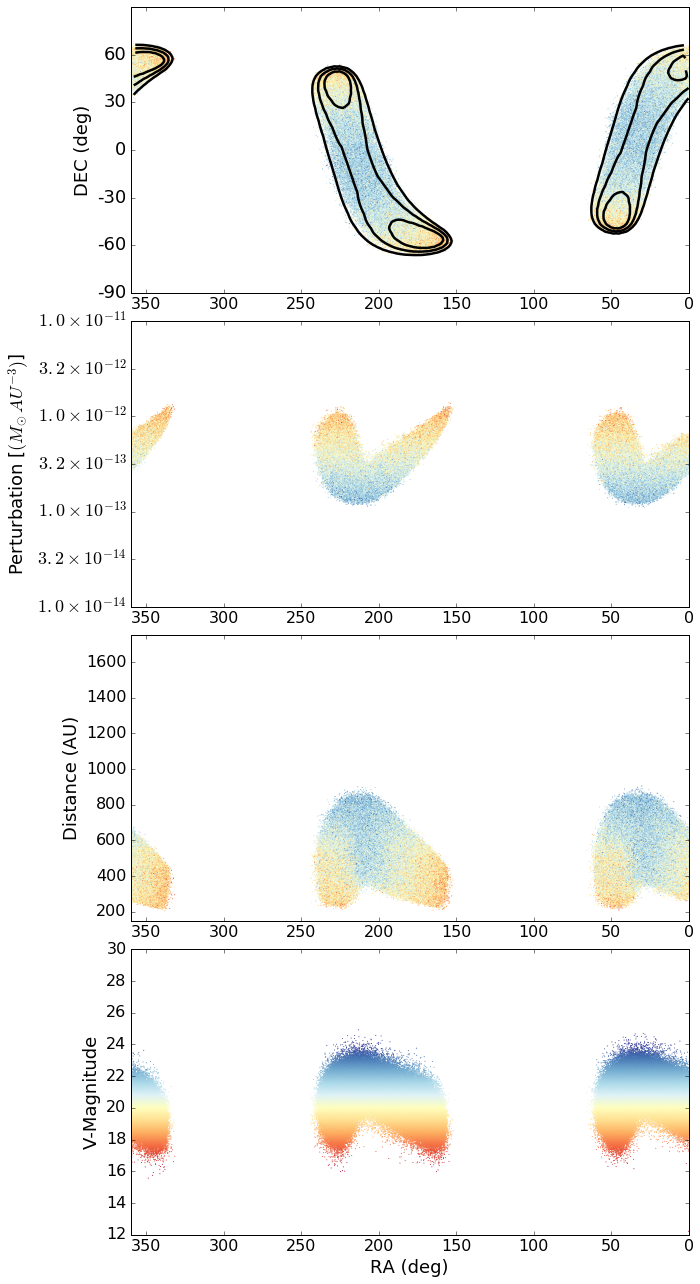}
    \caption{ %
    MCMC realization of preferred locations for \PX from Cassini-data alone. 
    {\bf Top: } RA and Dec. The $1,\,2,\,\&\,3\sigma$ contours from Figure \ref{FIG:SKY} are plotted in black. 
    {\bf 2nd-Top: } Perturbation magnitude as a function of RA (colored points).
    {\bf 2nd-Bottom: } Distance as a function of RA.
    {\bf Bottom: } V-band magnitude as a function of RA. 
    All points are colored according to their V-band magnitude in the bottom panel.
    The Cassini-data alone provide strong evidence that there is a tidal perturbation acting to modify the orbit of Saturn.
    The strength of the tidal perturbation is strongly confined to a range 
    $2\times10^{-13}\lsim\,M\,r^{-3} \lsim\, 10^{-12}\,M_{\odot}\,AU^{-3}$, or 
    $2.3\lsim\, \left(\frac{M}{10~M_{\oplus}}\right)\left(\frac{r}{700\,AU}\right)^{-3}\lsim\, 11.4$
    }
    \label{FIG:ALL_SKY_MCMC}
    \end{figure}

    To understand the detailed properties of objects with the sky-positions and the perturbation magnitudes of Figure \ref{FIG:SKY}, we use the MCMC approach described in Appendix \ref{APP:MCMC} to generate a distribution of objects with specific masses and orbital elements. We then use these to generate associated distributions of radii and magnitudes using the approach described in Appendix \ref{APP:MCMC:MAG}.
    We plot our results in Figure \ref{FIG:ALL_SKY_MCMC}. 

    %

    Our Cassini range-residual results clearly indicate a perturbing object occupying one of two narrow bands of the sky essentially perpendicular to the orbit of Saturn. 
    The favored perturbation scale is approximately $2\times10^{-13}\lsim\,M\,r^{-3} \lsim\, 10^{-12}\,M_{\odot}\,AU^{-3}$, or 
    \begin{equation}
        2.3 < \left(\frac{M}{10~M_{\oplus}}\right)\left(\frac{r}{700\,AU}\right)^{-3} < 11.4
        \label{EQN:PERT:SCALE}
    \end{equation}
    We emphasize that these results are completely independent of any other dynamical or observational constraints (e.g. \BBa\, and \BBB). 
    
    We note that our tidal model is symmetric, so by definition we will always find areas on opposing sides of the sky which have the same $\Delta\chi^2$.
    One might question whether we should try to break this degeneracy by using a moving-planet model. 
    We argue that (a) our experience in \citet{Holman2016} and the results of \citet{Hogg.1991} and \citet{Brunini.1992} show that the corrections are likely to be insignificant, and (b) the degeneracy can be broken by other methods (as described below).

\section{Results: Combined Constraints from Cassini and Dynamical Models}
\label{SECN:COMBINED}
We reiterate that our results presented in Section \ref{SECN:RESULTS:TIDAL} do \emph{not} depend at all on the dynamical results of \BBa\, and \BBB.
In this section, we combine the dynamical constraints from \BBa\, and \BBB\, with the Cassini ranging results. 

\subsection{Overlap} 
\label{SECN:COMBINED:OVERLAP} 
\BBa\, and \BBB\, perform a very large number of n-body simulations designed to narrow down the parameter space of orbits for \PX.
They do this by selecting those orbits for \PX which force a clustering in orbital alignment for ESDOs at least as tight as that seen in the observed Solar System objects. 
They find that they can roughly constrain the orbit of \PX to the following parameters:
\begin{itemize}
    \item Semi-major axis and eccentricity are restricted to a triangular region defined by the vertices ($a=300$~AU, $e=0.1$), ($a=300$~AU, $e=0.5$), and ($a=900$~AU, $e=0.8$).  
    \item Inclination is restricted to the range $i=22^{\circ}\,-\,40^{\circ}$.
    \item Argument of perihelion is restricted to the range $\omega=120^{\circ}\,-\,160^{\circ}$.
    \item ``Perihelion longitude'' is restricted to the range $226^{\circ}\,-\,256^{\circ}$.  We note that  \BBa~and \BBB\, refer to the ecliptic longitude of the point in the sky where the object appears at perihelion as the ``perihelion longitude'', not to be confused with the conventional orbital parameter called ``longitude of perihelion''.  
    \item There is \emph{no} constraint on mean anomaly.
    \item There is limited constraint on the mass: \BBB\, find that $10\,M_{\oplus}$ is acceptable, $0.1\,M_{\oplus}$ is ruled out and $1\,M_{\oplus}$ is disfavored. Given the uncertainties, we adopt a mass range $3-30\,M_{\oplus}$.
\end{itemize}

    \begin{figure}[thp]
    \centering
    \includegraphics[trim = 0mm 0mm 0mm 0mm, clip, angle=0, width=\columnwidth]{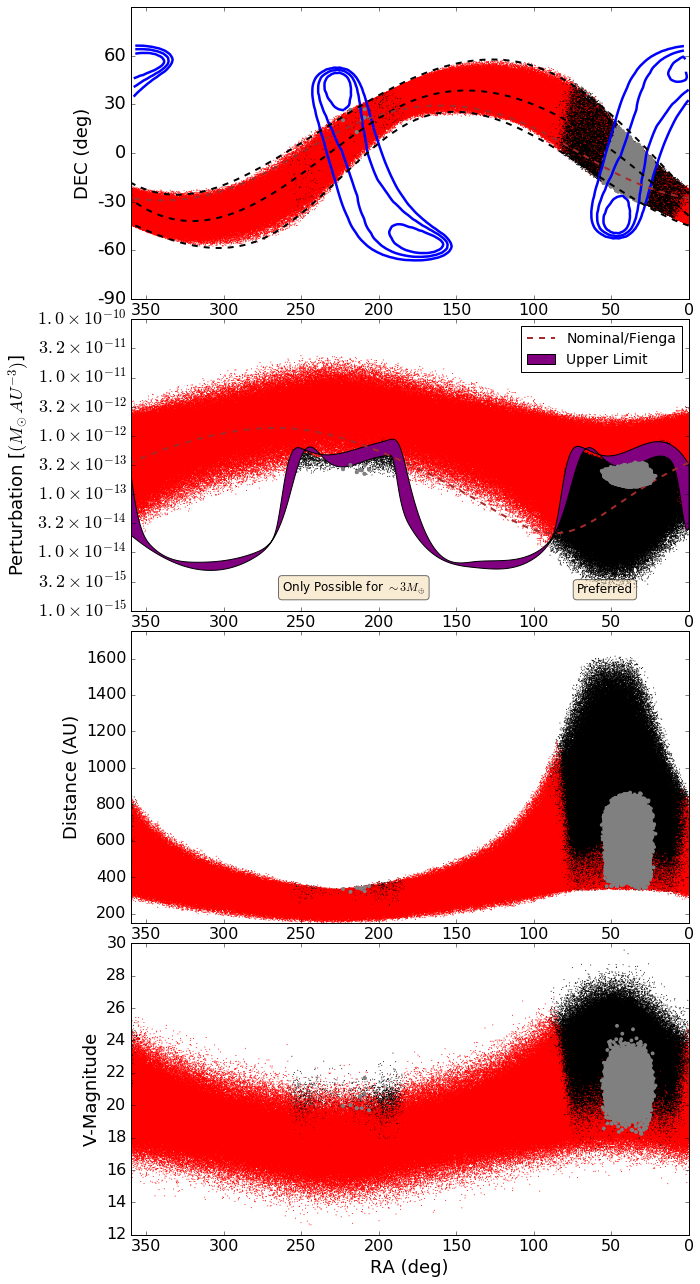}
    \caption{ %
    {\bf Top: } The range of RA and Dec occupied by the allowed orbits from \BBB. The $1,\,2,\,\&\,3\sigma$ contours from Figure \ref{FIG:SKY} are plotted in blue. 
    {\bf 2nd-Top: } Perturbation magnitude as a function of RA (points) and the upper-limits from Figure \ref{FIG:SKY} (purple).
    {\bf 2nd-Bottom: } Distance as a function of RA.
    {\bf Bottom: } V-band magnitude as a function of RA. 
    Our Cassini analysis rules out all of the points in red. 
    Only black \& gray points remain as possibilities, where the gray points are the subset of the black points that can cause a very significant improvement in fits to the Cassini data (within 3-sigma of the minimum). 
    A large fraction of the positions allowed by \BBB\, are ruled out completely.    
    Restricting the RA range to $5\arcdeg\lsim\,RA\lsim\,80\arcdeg$ offers a significant boost to the ``elimination'' plot of \BBB\, (their  Fig. 10) in which the unconstrained/unsearched region spans $30\arcdeg\lsim\,RA\lsim150\arcdeg$.
    }
    \label{FIG:SIMPLE_COMBO}
    \end{figure}
    %

Using this range of parameters, in Figure~\ref{FIG:SIMPLE_COMBO} we map out the possible locations on the sky that \PX could occupy (see also Figure~10 of \BBB). 
We sketched the same swath of allowed locations in the various panels of Figure \ref{FIG:SKY}.
The possible orbits of \PX from \BBB occupy an approximately $40\arcdeg$ wide band across the sky.
We also plot in Figure \ref{FIG:SIMPLE_COMBO} the path taken by the nominal orbit used by \F, illustrating that it is slightly offset from the median path of the allowed swath.

Further, we use the range of orbital parameters listed above and assuming a uniform, flat distribution in each parameter, randomly extract a value for each, as well as assigning a random mean anomaly for each set of parameters drawn. This defines an orbital location, sky position, and a tidal perturbation for a possible \PX.
We plot the declination, tidal parameter, distance and V-band magnitude as functions of RA in the various panels of Figure \ref{FIG:SIMPLE_COMBO}.
To calculate the V-band magnitude, we use the method described in Appendix \ref{APP:MCMC:MAG}. 

We then plot the upper limits on the tidal perturbations that we derived in Section \ref{SECN:RESULTS:TIDAL} (purple lines in the second panel of Figure \ref{FIG:SIMPLE_COMBO}).
We note that these are \emph{ranges} rather than single lines because we are plotting as a function of RA, and hence the results for a number of HEALPix tiles at different declinations are being combined. 

We use red to denote all of the points which are ruled out solely by our Cassini range analysis, where the criterion is whether the tidal perturbation is greater than the allowed upper limit in purple. 
The remaining, allowed points are plotted in black, with a subset of these being plotted in grey if they fall within the region that provides a significantly improved fit (within $3\sigma$ of the minimum).

We find that the Cassini data alone can rule out \PX existing almost anywhere in the range  $80\arcdeg\lsim\,RA\lsim\,365\arcdeg$, for the range of masses and orbital elements considered here.  
(By $80\arcdeg\lsim\,RA\lsim\,365\arcdeg$ we mean that the excluded region extends from RA=$80\arcdeg$ past $360\arcdeg$ to $5\arcdeg$.)
We refer the reader to Appendix \ref{APP:FIENGA} for a detailed consideration of why the constraints from \F~are not as restrictive as the ones we present. 

We note that the gray points, which provide a significantly improved fit (within $3\sigma$ of the minimum), are confined to a relatively small range of allowed tidal parameters.

Restricting the RA range to $5\arcdeg\lsim\,RA\lsim80\arcdeg$ substantially improves the ``elimination'' plot of \BBB\, (their  Fig. 10), in which the unconstrained/unsearched region of the sky spans $30^{\circ}<RA<150^{\circ}$.
Of particular value is the elimination of the region $90^{\circ}<RA<150^{\circ}$ through which the plane of galactic plane passes and for which it would be particularly challenging to search for a distant, faint, slow-moving planet.

We find that our analysis of the Cassini data can single-handedly exclude almost all proposed orbits from \BBB\, with RAs in the range $80\arcdeg\lsim\,RA\lsim\,365\arcdeg$.
There are a couple of tiny regions close to RA$=200\arcdeg$ and RA$=250\arcdeg$ in the bottom panel of Figure \ref{FIG:SIMPLE_COMBO} which we cannot exclude, but these would be ruled out by Pan-STARRS or other surveys~\citep{Brown.2016}.

Figure~\ref{FIG:SIMPLE_COMBO} illustrates that our Cassini analysis significantly constrains the range of orbits and masses proposed by \BBa\, and
\BBB.  Furthermore, Figure~\ref{FIG:SKY} demonstrates that 
the most highly significant regions from our Cassini analysis are outside the swath of orbits from \BBB.

In Section \ref{SECN:COMBINED:MCMC} we explore the dependence of the potential properties of \PX on the assumed orbital constraints from \BBB.  

In our discussion below, we consider further constraints on the location of \PX imposed by various observational surveys.

\subsection{Markov Chain Monte Carlo}
\label{SECN:COMBINED:MCMC}

In addition to the approach adopted in Section \ref{SECN:COMBINED:OVERLAP}, we also formally combine the dynamical constraints from \BBa\, with our all-sky Cassini ranging constraints using a Markov Chain Monte Carlo (MCMC) approach \citep[e.g.][]{Tegmark.2004,Ford.2005,Holman.2007}. 
Details of the implementation are provide in Appendix \ref{APP:MCMC}. 
In one set of runs we limit the allowed orbital parameters to the ranges specified by \BBB\, (and discussed above) and plot the results in green.
Then in a second set of runs we expand the allowed inclination range by from $22<i<40\arcdeg$ to  $12<i<50\arcdeg$ with the aim of allowing the routine to explore to lower declinations, coincident with the best-fit region of lowest $\chi^2$ from Figures \ref{FIG:SKY} and \ref{FIG:ALL_SKY_MCMC}. 
 
    \begin{figure}[thp]
    \centering
    \includegraphics[trim = 0mm 0mm 0mm 0mm, clip, angle=0, width=\columnwidth]{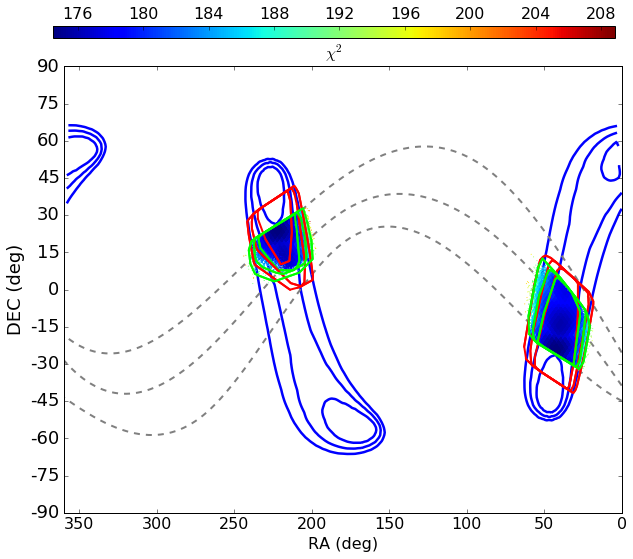}
    \caption{ %
    Results from our MCMC results, combining the dynamical constraints of \citet{Batygin.2016,Brown.2016} with our tidal constraints from Section \ref{SECN:RESULTS:TIDAL}.
    Results using priors based on the standard band of orbits from \BBB\, are plotted as green contours (enclosing $68.3\%$, $95.5\%$ and $99.7\%$ of the points respectively), while the equivalent contours for a \emph{wider} set of orbits are plotted in red.
    }
    \label{FIG:MCMC_A}
    \end{figure}
    %

Our MCMC results confirm and complement those of Figure \ref{FIG:SIMPLE_COMBO}.
\emph{Without} the prior constraints from \BBa and \BBB, we saw in Figure \ref{FIG:MCMC_A} that our Cassini range results favour thin strips of the sky running perpendicular to the orbit of Saturn. 
\emph{With} the priors imposed, we find in Figure \ref{FIG:MCMC_A} that the preferred regions on the sky occupy two distinct regions, close to 
(i) $(RA,DEC) \sim (40\arcdeg, -20\arcdeg)$, and 
(ii) $(RA,DEC) \sim (230\arcdeg, 20\arcdeg)$.
These regions are as close as possible to the minima of Figure \ref{FIG:SIMPLE_COMBO} while remaining within the swath of orbits suggested by the dynamics of \BBa\, and \BBB.

\subsection{Assessing the Results}
\label{SECN:ASSESS}

Our analysis is based on the Cassini range residuals with outliers rejected and with an updated estimate of the uncertainties.
It is critical that we assess the dependence of our results on that particular data set and uncertainty value.

To test the role of outliers, we repeat our analysis with the full Cassini data set, outliers included.  We do this with the two uncertainty values that have been mentioned before: the $\sigma=75$\,m level from \F~and the $\sigma=31.95$\,m level we used in our earlier analysis.  The fits are the same in the two cases; only the $\chi^2$ values are different.  However, the $\chi^2$ values affect the MCMC chains, as well as result that depends upon relative likelihood.
We provide some illustrative plots in Appendix \ref{APP:FIENGA}.

To further test the dependence upon our particular data set, we perform two different bootstrap analyses.  First, we use the most common form of bootstrap analysis, with involves generating new data sets by drawing an equal number of sample data points from the known data set, but {\it with replacement}.   After rejecting outliers, our original data set has 176 points.  We repeatedly draw 176 points from the original set, with replacement.  Although most points are drawn just once, some are drawn two or possibly three times, and others are skipped.  Each bootstrap data set is slightly different and thus yields different results.  The distribution of results gives a measure of their dependence on the specific data used.
We provide some illustrative plots in Appendix \ref{APP:BOOT}.
We find that the bootstrap analyses are largely similar to one another and to the nominal analysis.
The primary difference between the various bootstrap realizations is the location of the $1\sigma$ contour level, which in some iterations remains as two disconnected regions, while in others is joins into a single simply-connected region.

This approach is most sensitive to the role of individual data points, but it is less sensitive to the importance of sequences of data points that may be correlated.  We use a {\it block bootstrap} approach to investigate the possible role of such correlated groups.  Rather than drawing individual points from the original data set, we draw a sequence of consecutive points.  We already noted a systematic feature in the Cassini range residuals that spans most of 2013 and includes 10-20 data points.  We randomly draw 10 starting points within the data series and select 17 consecutive points starting from each of those.  If a sequence reaches the end of the data set, we continue that series at the beginning of the data set.  Finally, we draw six additional individual points to complete a sample of 176 points.  (We test the effect of varying the details of the block selection.)  
We provide some illustrative plots in Appendix \ref{APP:BOOT}.
We find that the location of the best-fit regions is insensitive to the {\it block bootstrap} reselection of data, although the \emph{depth} of the $\chi^2$ surface (and hence the statistical significance) does vary. 

As noted throughout, we are dependent upon the detailed \emph{unperturbed} ephemeris model of \F.  
Without their computationally intensive work to provide the unperturbed range-residuals, we would be able to proceed.
Similarly, if there are significant unmodeled effects or systematic errors in their work, then the conclusions we reach based upon their work will be invalidated.

\section{Discussion}
\label{SECN:DISC}

As we have stressed throughout this paper, while our simple, efficient model allows a broader parameter investigation than was possible for the numerically intensive study of \F, our work relies on there first existing an accurate, detailed ephemerides from the INPOP model of \F~(or JPL model of \citealt{Folkner.2014}) to provide a baseline set of Cassini range-residuals for the solar system with no \PX.
Likewise, if the reference ephemeris is inaccurate and the range residuals of the unperturbed model are in error, then our results will be invalidated.

In the best-fit, unperturbed model from \F~(the basis for all our investigations), it can be seen (in the top panel of Figure \ref{FIG:COMPARISON}) that there is a marked deformation in the shape of a coherent bowl-shaped depression around the 2013-2014 period. This short-timescale feature is not explained by tidal perturbations from \PX, either in our own work or that of \F. While we cannot explain such a feature within our model, we speculate that the deformation has a shape that we might expect to see if there was an unmodelled, (relatively) close-approach between a small perturbing body, and Saturn (or the Earth). The modelling of such perturbations is considered in \citet{Standish.2002} and is a standard part of large scale ephemerides projects \citep{Fienga.2011,Folkner.2014}.
The block-bootstrap analysis in Appendix \ref{APP:BOOT} demonstrates that our analysis is insensitive to this feature. 

In addition to the coherent trend between 2013-2014, the range residuals from the unperturbed model exhibit significant outliers, as illustrated in Figures \ref{FIG:COMPARISON} and \ref{FIG:HIST}. 
These outlying data points (which are retained in the analysis of \F) cause an overall degradation in the quality of the data and dilute the statistic certainly of the results. 
By excluding the outliers, we demonstrate (Figure \ref{FIG:COMPARISON2} and Appendix \ref{APP:FIENGA}) that we can increase the certainty with which we both exclude certain regions and favor other regions. 

Our Cassini range-residual results indicate a perturbing object occupying a narrow band of the sky essentially perpendicular to the orbit of Saturn with a favored perturbation scale 
$2\times10^{-13}<M\,r^{-3} < 10^{-12}\,M_{\odot}\,AU^{-3}$.
As tidal models are symmetric one might consider breaking this degeneracy by using a moving-planet model. 
However, the results of \citet{Hogg.1991}, \citet{Brunini.1992}, and \citet{Holman2016} show that the corrections will be insignificant for objects at the distance we find for \PX in Figures \ref{FIG:SIMPLE_COMBO}.
Moreover, the gross degeneracy across different sides of the sky is broken because the preferred orbits from \BBa\, and \BBB\, have a preferential alignment which places \PX  at smaller distances on one side of the sky.

\citet{Malhotra.2016} showed that if \PX has a period of $\sim11,411\,yrs$ (and therefore $a\sim506\,AU$), then its orbit would have integer period ratios with many of the ESDOs. Such a semi-major axis is completely consistent with the preferred semi-major axis from our results (see Figure \ref{FIG:SIMPLE_COMBO}).   Whether the preferred {\it sky location} for \PX that our analysis is also consistent with a resonant configuration remains to be tested.

 Although our analysis is based on the \F\, fits to the \emph{range} between the Cassini spacecraft and the Earth,
our models assume that the position of the Earth is well known, and that the orbit of Cassini around the Saturn system barycenter is also well known.  We verified these assumptions in section~\ref{SECN:VALIDATION}. The source of variation in the observed range stems from changes in the position of the barycenter of the Saturn system, rather than from  any uncertainty in the orbit of the Cassini spacecraft within the Saturn system.

As noted by \F, the constraints on \PX will improve with continued Cassini range measurements.  Furthermore, the approach we have presented here can be applied to other spacecraft ranging observations.  Extensive, highly precise range measurements to Mars-orbiting spacecraft have been collected and analyzed as part of programs to measure the gravity field of Mars.  Order of magnitude calculations suggest that \PX would exert a measurable influence on the Earth-Mars range, given the tidal parameter values we find from the Cassini data.  Although great care must be taken to account for the gravitation influence of the numerous asteroids that pass by Mars and Earth during the time span of the range observations~(Folkner, personal communication; \citealt{Kuchynka.2013}), we suggest analyzing those data sets with an eye toward more precisely constraining the location of \PX.

\begin{figure}[thp]
    \centering
    \includegraphics[trim = 0mm 0mm 0mm 0mm, clip, angle=0, width=\columnwidth]{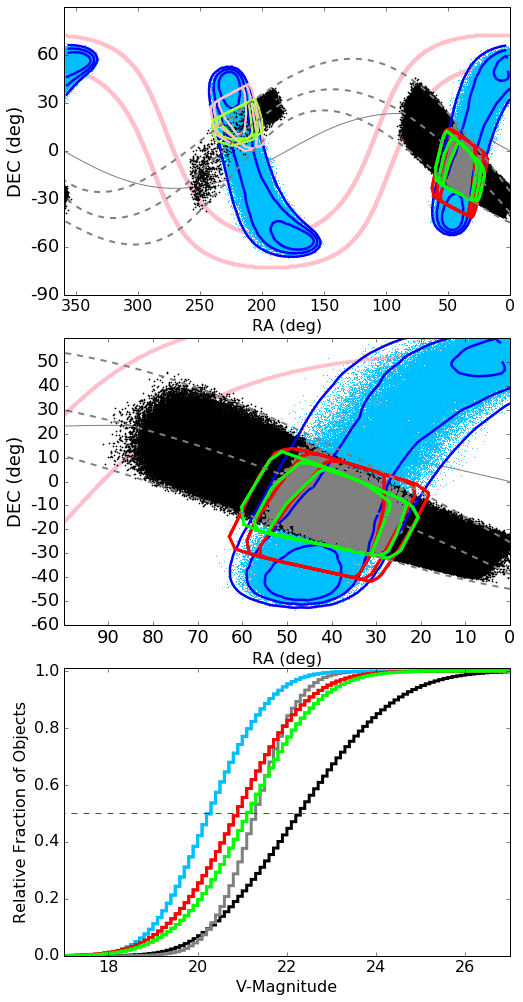}
    \caption{ %
    {\bf Top \& Middle: } Location.
    Dashed gray lines indicate the swath of orbits from \BBB.
    Pink lines show the galactic plane.
    Cyan line is the plane of the ecliptic.
    The contours from our Cassini analysis (Figure \ref{FIG:SKY}) are plotted in blue.
    Blue points are the all-sky MCMC results of Figure \ref{FIG:ALL_SKY_MCMC}.
    Black points are those from Fig. \ref{FIG:SIMPLE_COMBO} not ruled out by either our Cassini analysis or other observations.
    Grey points are the subset of the black points which would significantly improveme the Cassini data.
    The green contours are from the standard MCMC analysis of Section \ref{SECN:COMBINED:MCMC}.
    The red contours are from the broadened MCMC analysis of Section \ref{SECN:COMBINED:MCMC}.
    These patches are \emph{highly} localized to a region close to (RA,Dec)$=(40\arcdeg,-15\arcdeg)$
    It is clear that the preferred range of RA \& Dec is strongly restricted. 
    You should go and search there. 
    {\bf Bottom:} Magnitude distribution plotted in corresponding colors for the populations in the top two panels. 
    }
    \label{FIG:OBS}
    \end{figure}

In Figure \ref{FIG:OBS} we draw together the main results from our work. 
We illustrate the possible swath of orbits suggested by \BBB\, (dashed gray), the galactic plane (pink) and the plane of the ecliptic (solid gray).
We use black points to illustrate all those from Fig.\,\ref{FIG:SIMPLE_COMBO} which are not ruled out by our Cassini analysis.
We then use gray points to show the subset of the black points which would cause a highly significant improvement to the Cassini data (within $3\sigma$ of the absolute minimum).

We plot the $1,\,2,\&\,3\sigma$ contours from our Cassini analysis (Figure \ref{FIG:SKY}) in blue.
From this alone, it is already apparent that our best-fit results to the Cassini data, in conjunction with the broad swath of orbits from \BBB, strongly restrict the preferred parameter space to a region close to $RA, Dec\,\sim\,40\arcdeg, -15\arcdeg$, significantly away from the galactic plane.

We then overplot the results of the MCMC analysis in Section \ref{SECN:COMBINED:MCMC} using green contours for the analysis based on the orbital constraints of \BBB, and then red contours for an analysis in which the orbital constraints of \BBB~are loosened by $10$ degrees. 
For all of the plotted distributions, we provide the distribution of expected V-band magnitudes. 

Figure~\ref{FIG:OBS} shows the predicted location of \PX,  a highly confined region of sky defined by the intersection of our Cassini results with the dynamical constraints from \BBa\, and \BBB.  
We suggest the following observational strategy: 
\begin{itemize}
    \item Begin by searching the region within the $1\,\sigma$ contour (inner green line) of Figure \ref{FIG:OBS}, the region that results from combining our analysis of the Cassini data with the dynamical constraints of \BBa\, and \BBB; 
    \item Next, broaden the search region radially from that region by expanding into (i) the region within the $2\sigma$ green contour, and (ii) the $1\,\&\,2\sigma$ regions from the \emph{broader} MCMC analysis (red contours from Figure \ref{FIG:OBS}) which stretch below the swath of orbits from \BBB. 
    \item Then move out to the regions within the $3\sigma$ green and red contours of Figure \ref{FIG:OBS}, and then perhaps also the lower blue $1\sigma$ contour Figure \ref{FIG:OBS} if one is willing to stray that far from the lower boundary of the swath preferred by \BBB. 
\end{itemize}

    %

We note that many of the best-fit solutions of Figure \ref{FIG:OBS} are \emph{bright}, with V-band magnitudes spanning the range $18^{\rm th}-26^{\rm th}$ magnitude. 
We acknowledge that many of these could be ruled out by previous observational surveys, as discussed by \BBB. 
However, we are somewhat more conservative than \BBB\, regarding the completion limits of the available observational surveys, particularly in the regions of the southern sky around (RA, Dec) $=\,(40\arcdeg, -15\arcdeg)$.  In appendix~\ref{APP:FAINT} we present an alternative version of Figure~\ref{FIG:OBS} that excludes solutions that are brighter than $V=22.5$.

\section{Conclusions}
\label{SECN:CONC}
We significantly constrain the sky position, distance, and mass of a possible additional, distant planet in the solar system by examining its influence on the distance between Earth and the Cassini Spacecraft.

Without reference to any other investigation,  the Cassini data alone strongly suggest a distant planet, with tidal parameter in the range $1\times10^{-13}\,-\,1\times10^{-12}\,M_{\odot}\,AU^{-3}$, is measurably influencing the orbit of Saturn.
The location of this planet, again from the Cassini data alone, is confined to one of two thin strips, on opposite sides of the sky, that run perpendicular to the orbit of Saturn.

When we combine our constraints from the Cassini data with the dynamical constraints from \citet{Batygin.2016} and \citet{Brown.2016}, our preferred region is centered approximately at (RA, Dec) $=\,(40\arcdeg ,-15\arcdeg)$, and extends $\sim20\arcdeg$ in all directions.
We recommend this region as being of the highest priority for an efficient observational campaign.


\section*{Acknowledgments}

We thank William Folkner, Eric Agol, Tsevi Mazeh, Norman Murray, Pedro Lacerda, Wesley Fraser, and Joshua Winn for helpful discussions.

MJH and MJP gratefully knowledge
NASA Origins of Solar Systems Program grant NNX13A124G, 
NASA Origins of Solar Systems Program grant NNX10AH40G via sub-award agreement 1312645088477, 
BSF Grant Number 2012384,
NASA Solar System Observations grant NNX16AD69G, 
as well as support from the Smithsonian 2015 CGPS/Pell Grant program.

\bibliography{references}


\appendix

\newpage

\section{Excluding Objects Brighter than $22.5$} 
\label{APP:FAINT}
\begin{figure}[thp]
    \centering
    \includegraphics[trim = 0mm 0mm 0mm 0mm, clip, angle=0, width=0.6\columnwidth]{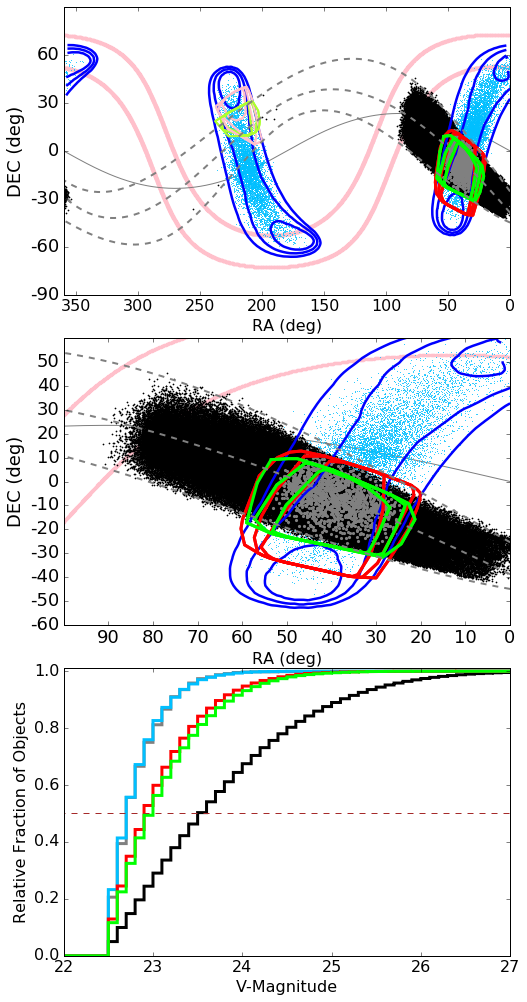}
    \caption{ %
    Alternative version of Figure \ref{FIG:OBS}, in which only points fainter than V magnitude 22.5 have been plotted. 
    It is still the case that all populations have a significant number of members remaining within the preferred region of the sky.
    The most significant reduction occurs for the points within the blue $1\sigma$ region, suggesting that if objects brighter than magnitude 22.5 \emph{are} reliably ruled out, then there is less reason search far from the preferred swath of \BBB. 
    }
    \label{FIG:OBS:ALT}
    \end{figure}

In Figure \ref{FIG:OBS} of section \ref{SECN:DISC}, we presented a summary of our preferred region on the sky to search for \PX.
In Figure \ref{FIG:OBS} we plotted preferred solutions without constraining their visual magnitude. 
As discussed by \BBB, there are significant constraints on the possible magnitude \PX could have and still be consistent with previously conducted searches.
While we are somewhat more pessimistic than \BBB\, in our beliefs regarding the magnitude limits that have realistically been achieved in the southern sky, we acknowledge that reasonable bounds do exist. Hence, in Figure \ref{FIG:OBS:ALT} we present an alternative version of Figure \ref{FIG:OBS} in which we plot \emph{only} those realizations for \PX in which the magnitude is fainter than 22.5.
We find that it is still the case that all populations have a significant number of members remaining within the preferred region of the sky.
The most significant reduction occurs for the points within the blue $1\sigma$ region, suggesting that if objects brighter than magnitude 22.5 \emph{are} reliably ruled out, then there is less reason to search from the preferred swath of \BBB.

    %

\newpage
\section{Further Details on Reproducing the Results of \citet{Fienga.2016}} 
\label{APP:FIENGA}
\subsection{Mislabeled True Anomaly}
\begin{figure*}[thp]
    \begin{minipage}[b]{\textwidth}
    \centering
    \includegraphics[trim = 0mm 0mm 0mm 0mm, clip, angle=0, width=0.45\textwidth]{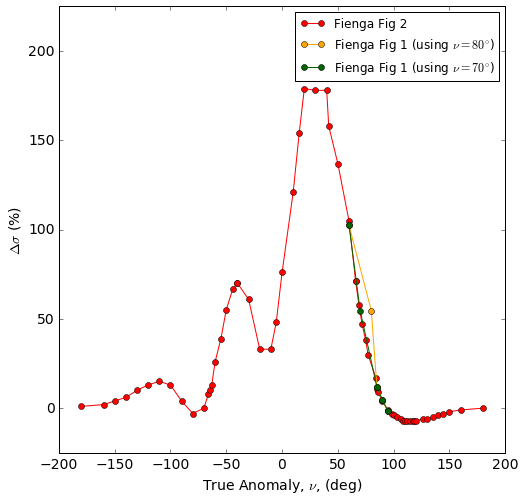}
    \caption{ %
    Comparison between the $\Delta\sigma$ measure calculated from data extracted from Figure (1) of \citet{Fienga.2016} (green and orange data), compared to the $\Delta\sigma$ measure extracted directly from Figure (2) of \citet{Fienga.2016} (red). The measures are inconsistent if one assumes the value of  $\nu=80\arcdeg$ (orange) written in their manuscript, but the measures become consistent if one assumes a typo, and that the real value is in fact $\nu=70\arcdeg$ (green data). 
    }
    \label{FIG:APP:F1F2}
    \end{minipage}
    \end{figure*}

We noted in Section \ref{SECN:VALIDATION} and Figure \ref{FIG:COMPARISON} that we believe the second panel of Figure (1) in \citet{Fienga.2016} is mislabelled. We verify this in Figure \ref{FIG:APP:F1F2} where we plot a comparison between the $\Delta\sigma$ measure calculated from data extracted from Figure (1) of \citet{Fienga.2016}, and compare it to the $\Delta\sigma$ measure extracted directly from Figure (2) of \citet{Fienga.2016}. We show that the measures are inconsistent if one assumes the value of  $\nu=80\arcdeg$ written in their manuscript, but that the measures become consistent if one assumes a typo, and that the real value is in fact $\nu=70\arcdeg$. This difference is \emph{un}important for the rest of our analysis, as all of our work is based on their \emph{unperturbed} data set.

    %

\subsection{$\chi^2$ versus $\Delta\sigma$}
\label{APP:FIENGA:CHI}
As we discussed in Section \ref{SECN:VALIDATION}, \F~use a somewhat unusual statistical measure to evaluate the significance of their perturbed fits to the Cassini data.
They state that their criteria for \emph{rejecting} perturbation models is  a relative increase in the post-fit residuals of more than $10\%$ after the addition of \PX. 
Similarly they accept improved fits as significant if they result in a relative decrease in the post-fit residuals of more than $6\%$ after the addition of \PX. 

Using $\Delta\chi^2=N\left(\Delta\sigma^2+2*\Delta\sigma\right)$, we calculate that a $10\%$ increase  in $\Delta\sigma$ corresponds to $\Delta\chi^2\sim\,40$, a rather larger figure than our own $\Delta\chi^2_{3\sigma}\sim9$ for a 1-D.o.F model. Similarly, $\Delta\sigma$ of $-6\%$ corresponds to $\Delta\chi^2\sim\,-22$, forcing their results to be rather conservative with regards to both the regions it can exclude as well as the regions it classifies as giving improved fits. 

\begin{figure*}[thp]
    \begin{minipage}[b]{\textwidth}
    \centering
    \includegraphics[trim = 0mm 0mm 0mm 0mm, clip, angle=0, width=0.45\textwidth]{LINE}
    \includegraphics[trim = 0mm 0mm 0mm 0mm, clip, angle=0, width=0.45\textwidth]{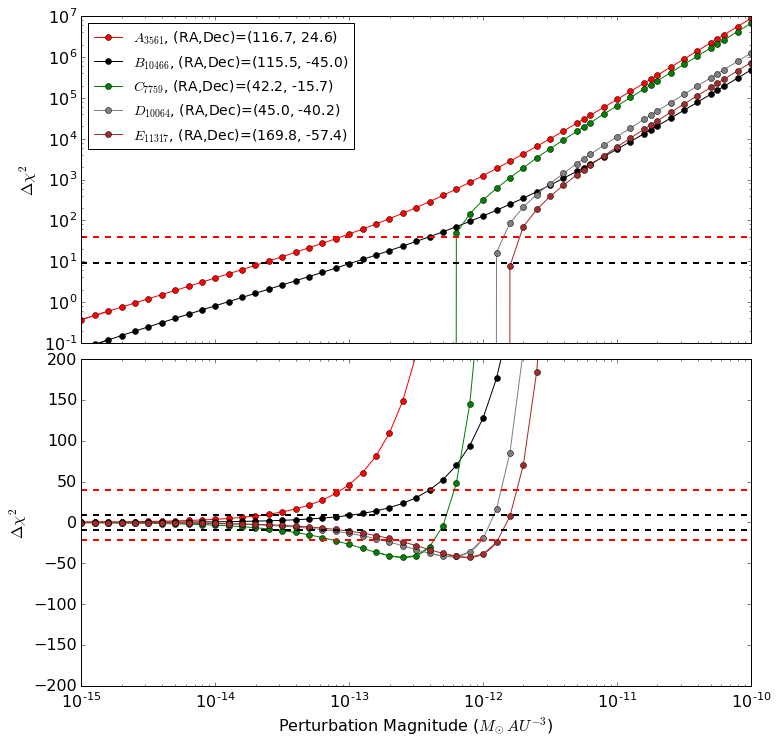}\\
    \caption{ %
    Comparison between Figure \ref{FIG:LINE} (left) and version including all outlying data points (right).
    The minima become significantly less deep and hence the range of ``improving'' perturbations also becomes significantly narrower and skews to higher tidal parameters. 
    }
    \label{FIG:APP:LINE}
    \end{minipage}
    \end{figure*}

\begin{figure*}[thp]
    \begin{minipage}[b]{\textwidth}
    \centering
    \includegraphics[trim = 0mm 0mm 0mm 0mm, clip, angle=0, width=0.33\textwidth]{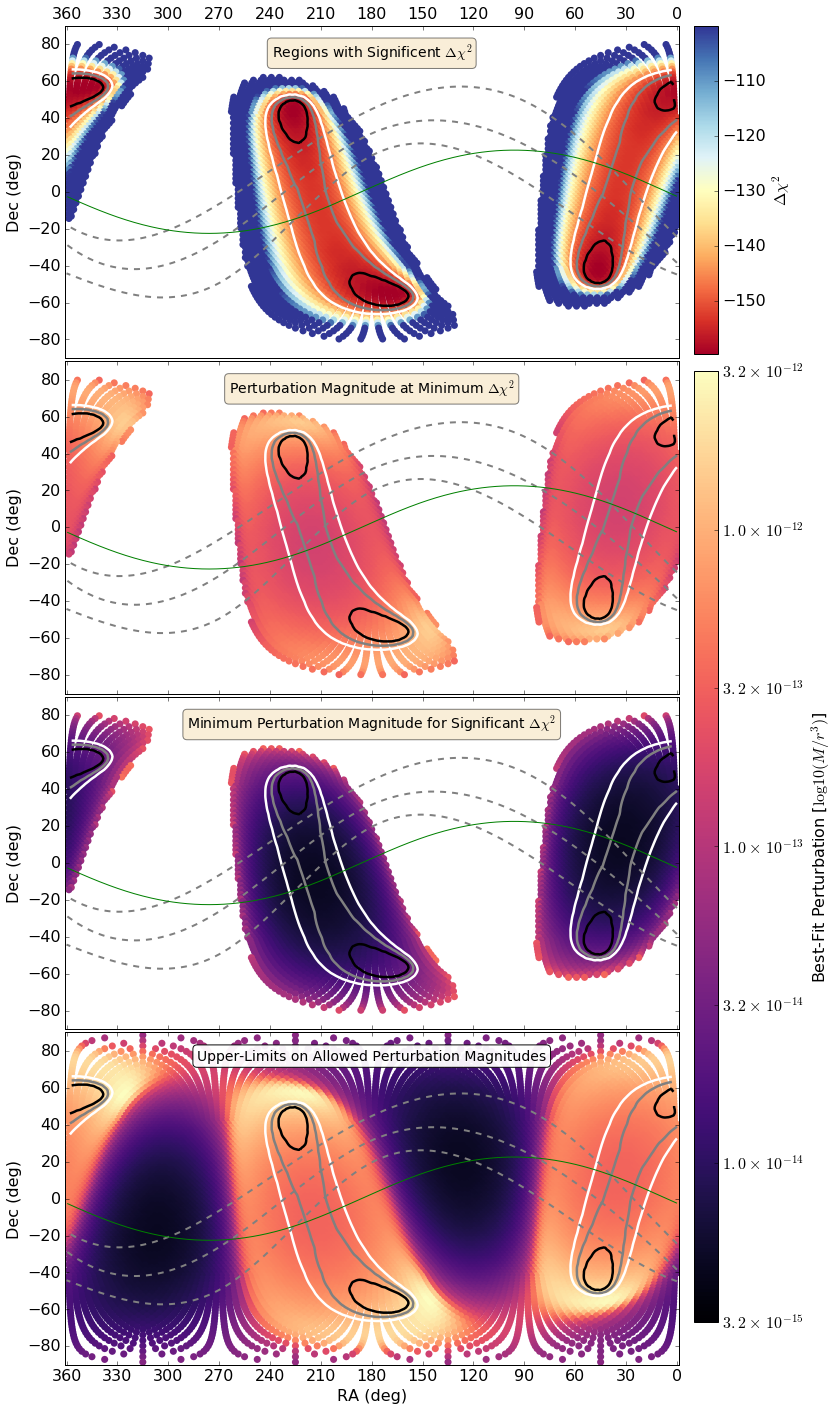}
    \includegraphics[trim = 0mm 0mm 0mm 0mm, clip, angle=0, width=0.33\textwidth]{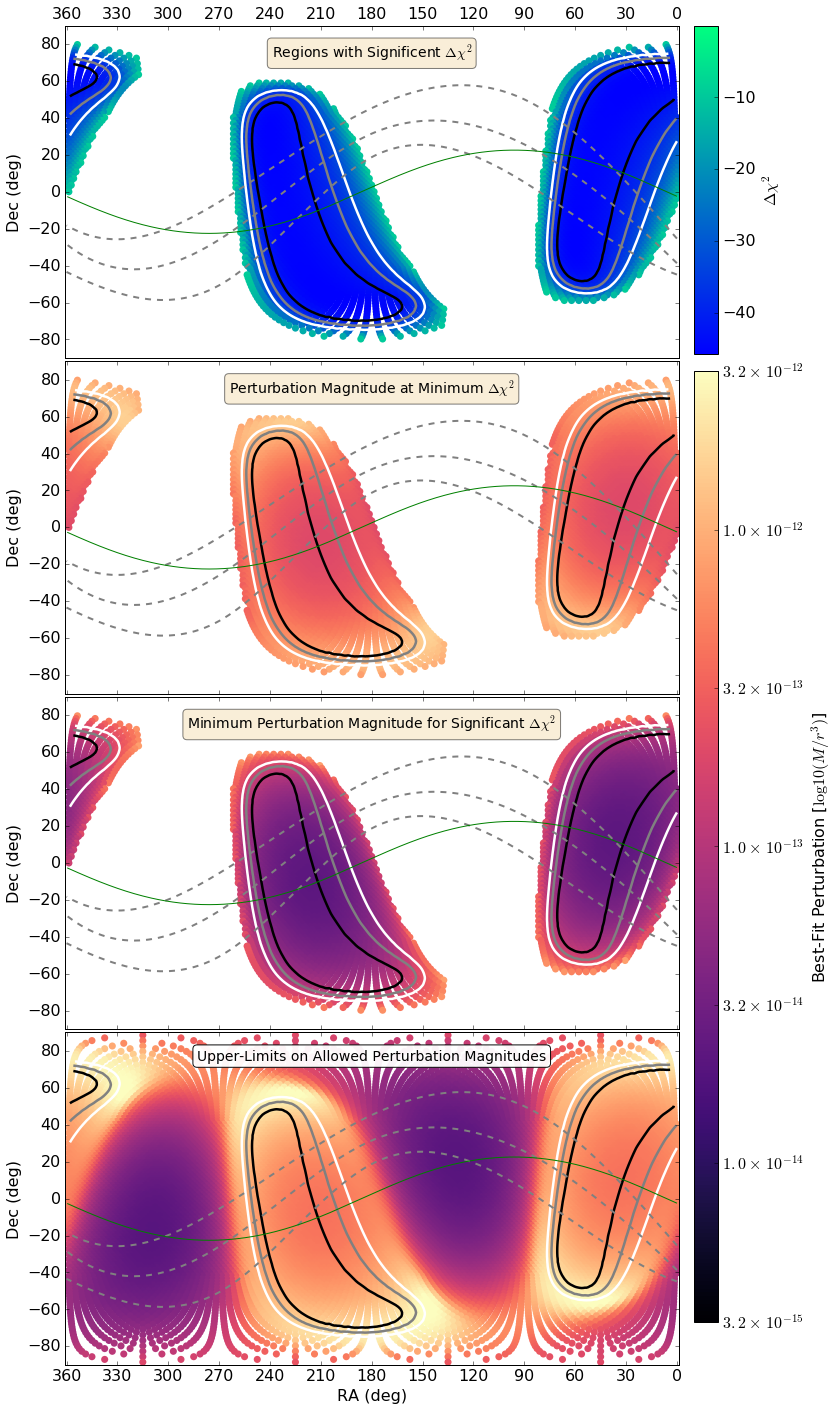}
    \includegraphics[trim = 0mm 0mm 0mm 0mm, clip, angle=0, width=0.33\textwidth]{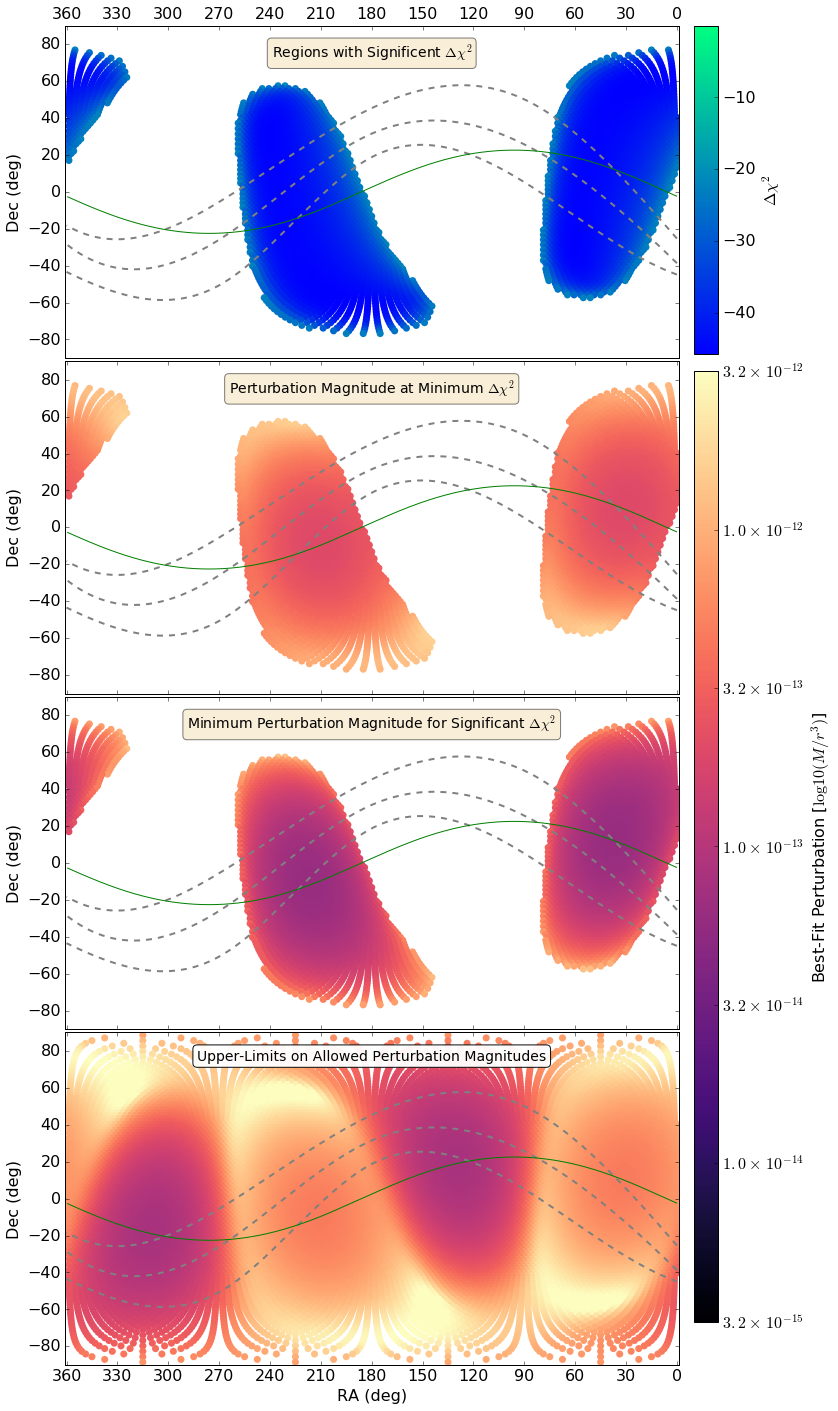}\\
    \caption{ %
    Comparison between Figure \ref{FIG:SKY} (left), version including all outlying data points (middle), and a version using all data points and making cuts based on the \citet{Fienga.2016} $\Delta\sigma$ criteria of $+10\%$ and $-6\%$ as discussed in Appendix \ref{APP:FIENGA:CHI} (right).
    Using all data (including outliers, middle panel) and expanding the assumed errors to $\sim75$m, the minima become significantly shallower and the $1/2/3\sigma$ contours expand outwards to encompass a larger area. In the bottom panels, the upper-limits that we can rule out become higher, yielding less robust exclusion of perturbers over large areas of the sky.
    When we use the $\Delta\sigma$ criteria of \citet{Fienga.2016} (right), we find the key difference becomes the inability to rule out small tidal perturbations across large portions of the sky. This is why \citet{Fienga.2016} were unable to decisively rule out \PX for large ranges of true anomaly.
    }
    \label{FIG:APP:SKY}
    \end{minipage}
    \end{figure*}

In Figures \ref{FIG:APP:LINE} and \ref{FIG:APP:SKY} we repeat our entire analysis using data sets that include all outlier data points that we reject in our primary analysis.  Figure~\ref{FIG:APP:SKY} illustrates two primary results.  First, including the outlier data points results in less statistical significance in the overall results.  Second, using the statistical measure \F\, further dilutes the strength of the conclusions that one can reach.

    %

    %

\newpage

\section{MCMC Methodology, Orbital Location and Visible Magnitudes}
\label{APP:MCMC}
%
%
In Sections \ref{SECN:CASS:SUMM} and \ref{SECN:COMBINED:MCMC} we illustrated results obtained using Markov Chain Monte Carlo (MCMC) techniques \citep[e.g.][]{Tegmark.2004,Ford.2005,Holman.2007} to generate distributions of masses and orbital elements for \PX that satisfy various constraints. 
We document here the pertinent technical details.

In general, given an initial set of mass and orbital parameters for \PX and $\chi^2_0$ (the current value), we first select a possible subsequent set by sampling from the distribution of orbits that satisfy the dynamical constraints.  Specifically, we choose one of the parameters  ($GM_p$, $a$, $e$, $i$, $\Omega$, $\omega$, $M$) at random, and we add to it a small offset drawn randomly from fixed gaussian distributions.  We then test if the new orbital parameters are still within the allowed limits.  If not, the system stays in the same state.  If the new orbital parameters satisfy the dynamical constraints, we evaluate the new $\chi^2$, based on our fits to the Cassini data.  If $\Delta\chi^2=\chi^2-\chi^2_0<0$, the jump is accepted.  Otherwise the jump is accepted with probability $\propto\exp(-\frac{1}{2}\Delta\chi^2)$.
The specific limits used as priors in the different sections are detailed below. 

For all MCMC runs, irrespective of the specific limits being implemented we abopt the following methodology:
We start 20 independent chains.
We run the chains for $10^8$ iterations.
We test for convergence and mixing \citep{Gelman.1992}.
We reject the first $50\%$ of each chain.

For many/all of the MCMC realizations we go on to calculate radii and then visible magnitudes for each of the  masses and positions generated by the MCMC chains. 
Further details of the process of assigning a radius and magnitude are provide in Appendix \ref{APP:MCMC:MAG} below. 
We note that we do \emph{not} include either radii or observational (magnitude) constraints as MCMC priors.

\subsubsubsection{MCMC For Cassini-Only Data }
\label{APP:MCMC:CASS}
In Section \ref{SECN:CASS:SUMM} we used an MCMC method to select masses and orbital elements for data in which the sky-plane position and perturbation magnitudes are constrained \emph{solely} by the Cassini data. 
In this realization we constrain the semi-major axes, eccentricities and masses to the same ranges as described in Appendix \ref{APP:MCMC:COMB} for the realizations constrained by the \BBB\, orbits.
However, we let all angular elements (inclination, longitude of ascending node, argument of pericenter, and mean anomaly) range freely, meaning that there is \emph{no} preferred orientation on the sky. 

\subsubsubsection{MCMC For Combinations of Cassini Data with Dynamical Constraints from \BBB}
\label{APP:MCMC:COMB}
In Section \ref{SECN:CASS:SUMM} we used an MCMC method to select masses and orbital elements for data in which the sky-plane position and perturbation magnitudes are jointly constrained by the Cassini data and by the preferred orbits of \BBB. 
These orbits are confined to the following parameter space:
\begin{itemize}
    \item Semi-major axis and eccentricity are restricted to a triangular region defined by the vertices ($a=300$~AU, $e=0.1$), ($a=300$~AU, $e=0.5$), and ($a=900$~AU, $e=0.8$).  
    \item Inclination is restricted to the range $i=22^{\circ}\,-\,40^{\circ}$.
    \item Argument of perihelion is restricted to the range $\omega=120^{\circ}\,-\,160^{\circ}$.
    \item ``Perihelion longitude'' is restricted to the range $226^{\circ}\,-\,256^{\circ}$.
    \item There is \emph{no} constraint on mean anomaly.
    \item There is limited constraint on the mass: \BBB\, find that $10\,M_{\oplus}$ is acceptable, $0.1\,M_{\oplus}$ is ruled out and $1\,M_{\oplus}$ is disfavored. Given the uncertainties, we adopt a mass range $3-30\,M_{\oplus}$.
\end{itemize}

A number of patterns related to these two regions occur in the output distributions:
\begin{itemize}
    \item The preferred masses split bi-modally between the two regions, with the preferred mass for the $(RA,DEC) \sim (40\arcdeg, -20\arcdeg)$ being strongly pushed up against the imposed $30M_{\oplus}$ upper-limit, while for the  $(RA,DEC) \sim (230\arcdeg, 20\arcdeg)$ region it strongly pushes down to the lower-limit of $3M_{\oplus}$.
    \item If the object is at $(RA,DEC) \sim (230\arcdeg, 20\arcdeg)$, then it is \emph{close}: $r\sim350\,AU$, while if it is at $(RA,DEC) \sim (40\arcdeg, -20\arcdeg)$ it is much more distant: $r\sim700\,AU$.
    \item The tidal parameter is well confined to a value 
        $\sim 2.3^{+0.35}_{-0.39}\times10^{-13}$; 
    \item As expected for a tidal model, the mass and distances are essentially degenerate. 
    \item Irrespective of region on the sky, the semi-major of the preferred orbit is well confined  to a single (but broad) peak, $a=499^{+72}_{-85}$. We note that this is completely consistent with the value suggested by \citet{Malhotra.2016} based on the argument that $a\approx\,506$ would place \PX in multiple mean motion resonances with the various Sedna-like objects. 
    \item Irrespective of region on the sky, the eccentricity of the preferred orbit is well confined  to a single peak, $e=0.43^{+0.10}_{-0.10}$
\end{itemize}

\subsubsubsection{MCMC For Combinations of Cassini Data with \emph{Relaxed} Dynamical Constraints from \BBB.}
\label{APP:MCMC:RELAX}
In Section \ref{SECN:COMBINED:MCMC} we used an MCMC method to select masses and orbital elements for data in which the sky-plane position and perturbation magnitudes are jointly constrained by the Cassini data and by a \emph{relaxed} version of the preferred orbits from \BBB. 
For this realization we kept the same constraints on the semi-major axes, eccentricities, argument of perihelion, perihelion longitude, and masses as detailed in Appendix \ref{APP:MCMC:COMB}.
However, we relax the inclination constraint such that inclination is restricted to the range $i=12^{\circ}\,-\,50^{\circ}$.

\newpage
\subsection{Assigning a Radius and Magnitude}
\label{APP:MCMC:MAG}
To calculate the V-band magnitude, we use the ``forecaster'' routine of \citet{Chen.2016} (which is conditioned on a sample of 316 exoplanets with well-constrained masses and radii) to select a sample radius for a given planetary mass.
Because our masses are in the range $3-30M_{\oplus}$, $>95\%$ of our samples are classed as ``Neptunian'', with the rest being classified as ``Earth-like''. 
For those objects classified as ``Neptunian'' we follow the simple prescription of \BBB and assume an albedo of~$0.3$.
For those few classified as Earth-like, given the vast albedo range observed for TNOs \citep{SantosSanz.2012}, one could justify almost any choice of albedo. 
For simplicity, we again adopt~$0.3$.
If \PX is of radius, $R$, and at a heliocentric distance $r$, we can use the approximate relation $V=-26.74 - 5\times\log_{10}{\left( R \times 0.3^{0.5} / ( r(r-1)) \right)}$ to convert to V-band magnitude.

\newpage
\section{Appendix: Bootstrap} 
\label{APP:BOOT}
\begin{figure*}[thp]
    \begin{minipage}[b]{\textwidth}
    \centering
    \includegraphics[trim = 0mm 370mm 0mm 0mm, clip, angle=0, width=0.49\textwidth]{ALL_SKY_BEST}
    \\
    \includegraphics[trim = 0mm 370mm 0mm 0mm, clip, angle=0, width=0.49\textwidth]{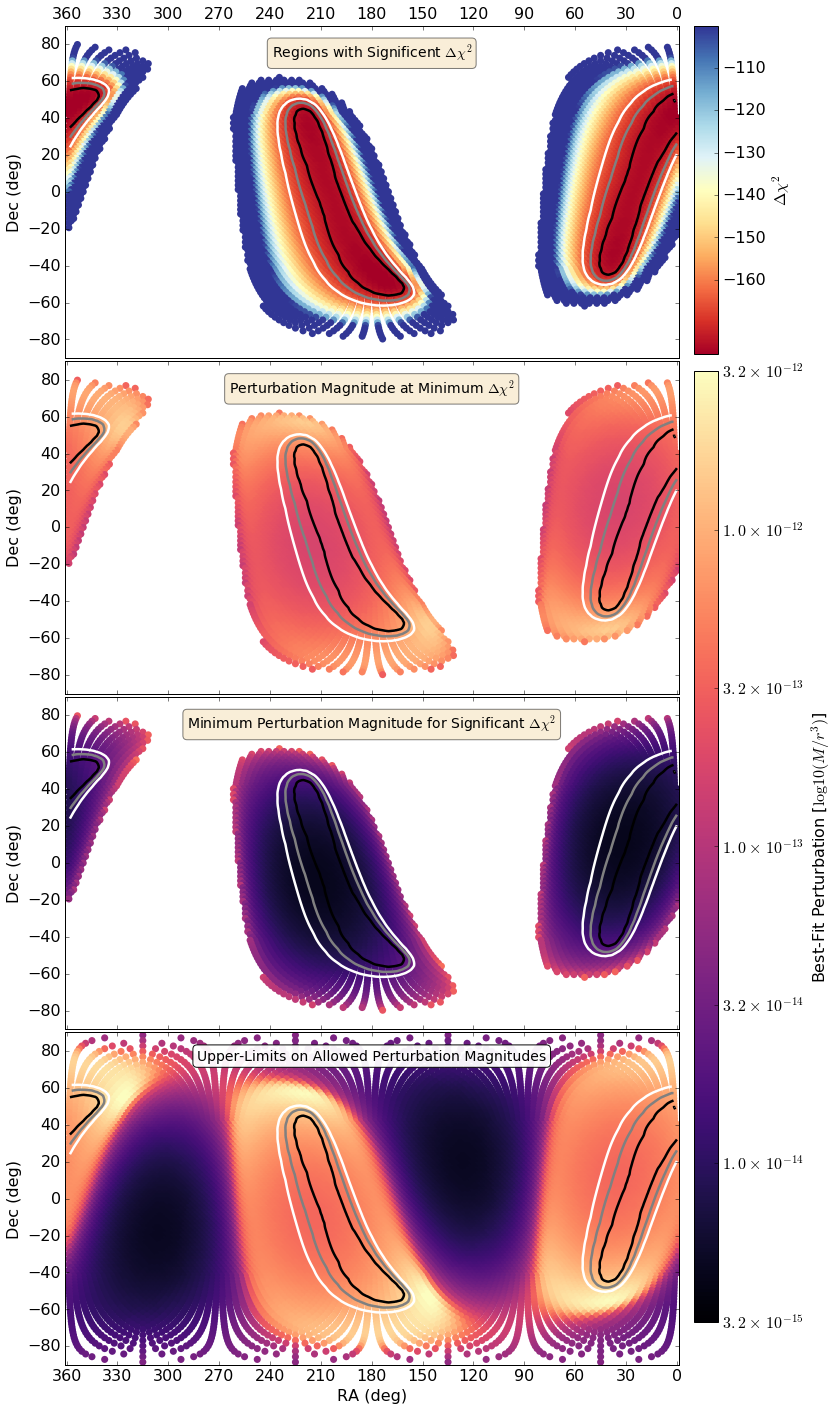}
    \includegraphics[trim = 0mm 370mm 0mm 0mm, clip, angle=0, width=0.49\textwidth]{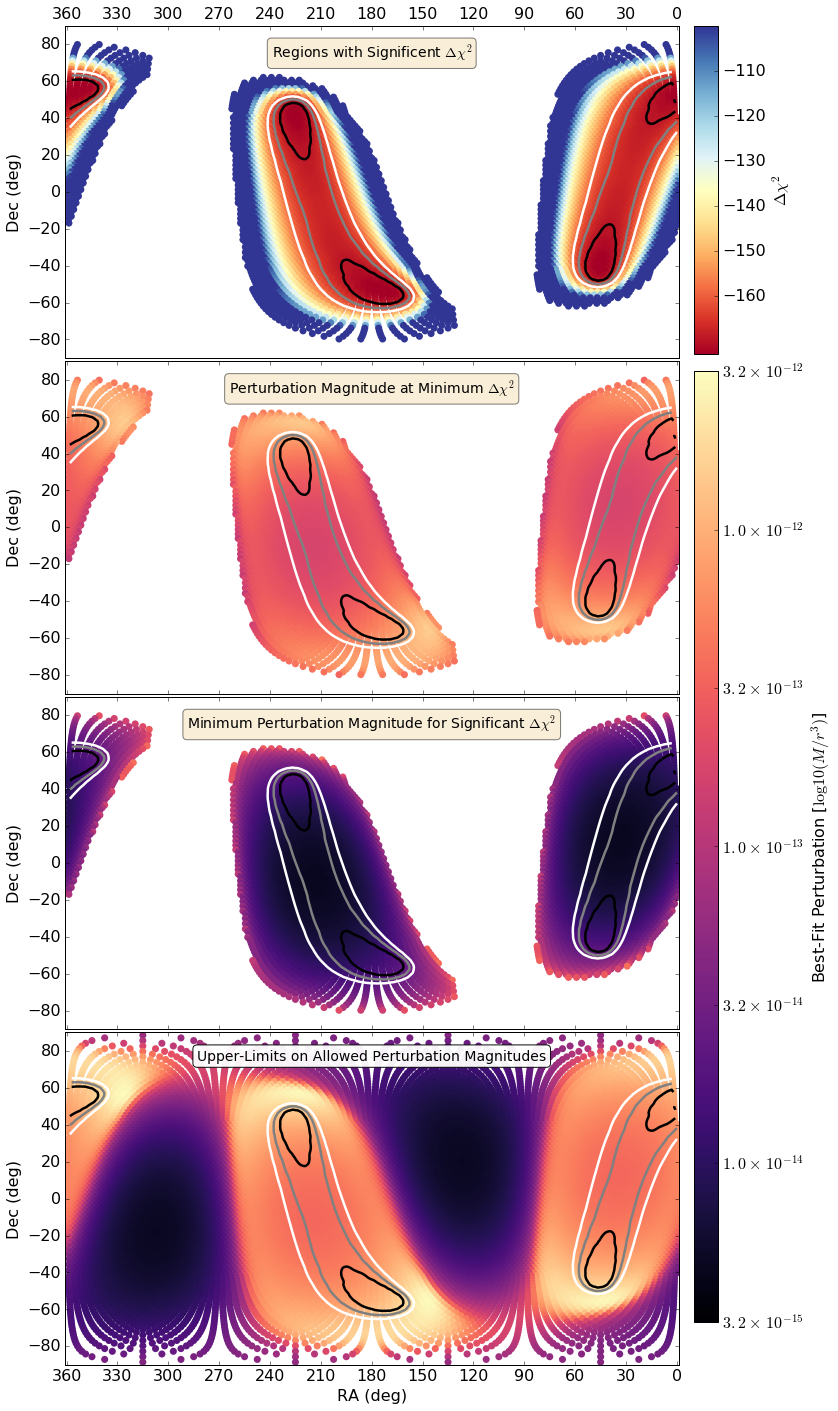}
    \\
    \includegraphics[trim = 0mm 370mm 0mm 0mm, clip, angle=0, width=0.49\textwidth]{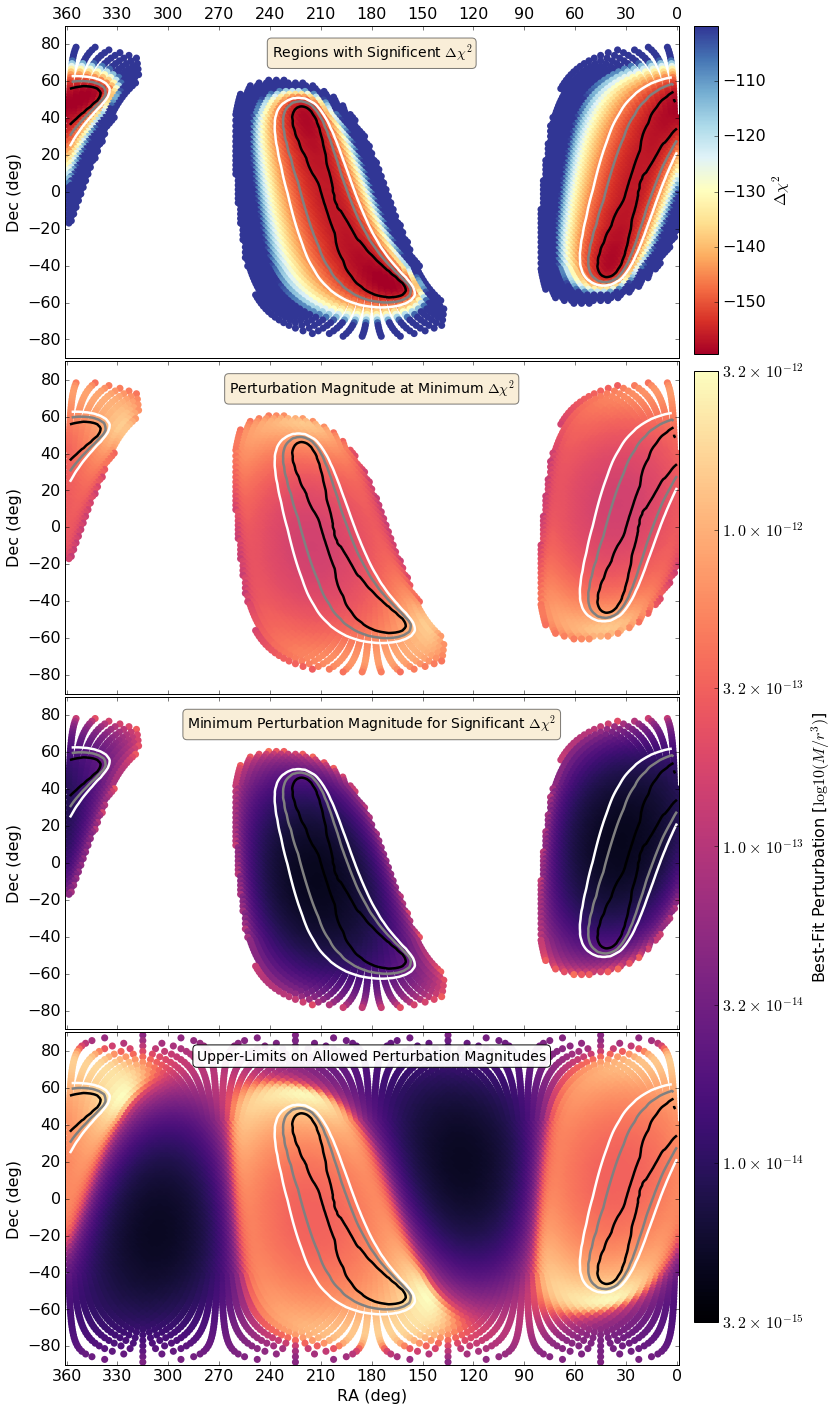}
    \includegraphics[trim = 0mm 370mm 0mm 0mm, clip, angle=0, width=0.49\textwidth]{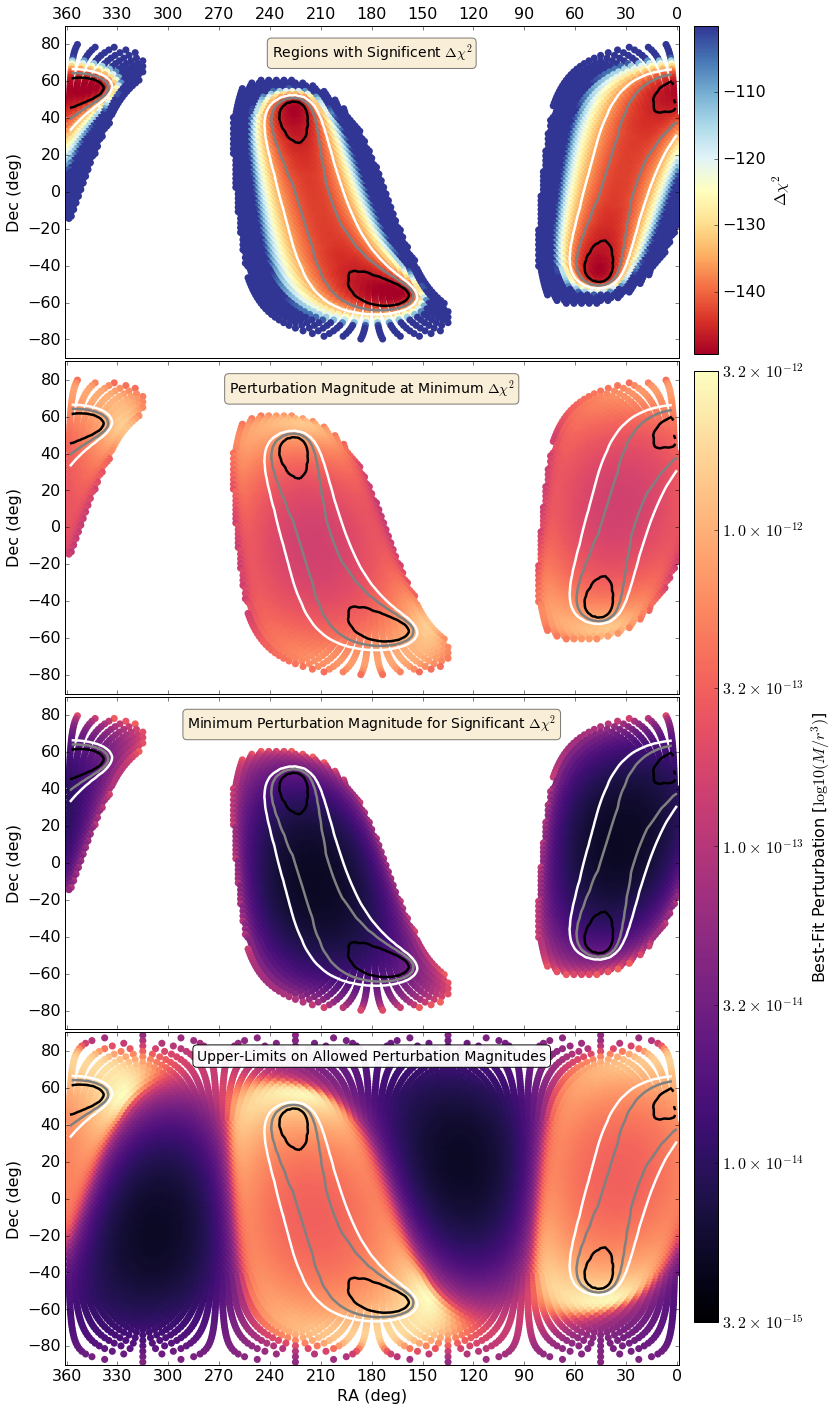}
    \caption{ %
    {\bf Top: } Standard results from Figure \ref{FIG:SKY}
    {\bf Others: } Results from four different bootstrap realizations (outliers rejected). 
    The overall results look broadly similar for the different bootstrap iterations, with the main difference being in the location of the $1\sigma$ contour. 
    }
    \label{FIG:APP:BOOT}
    \end{minipage}
    \end{figure*}

\begin{figure*}[thp]
    \begin{minipage}[b]{\textwidth}
    \centering
    \includegraphics[trim = 0mm 370mm 0mm 0mm, clip, angle=0, width=0.49\textwidth]{ALL_SKY_BEST}
    \\
    \includegraphics[trim = 0mm 370mm 0mm 0mm, clip, angle=0, width=0.49\textwidth]{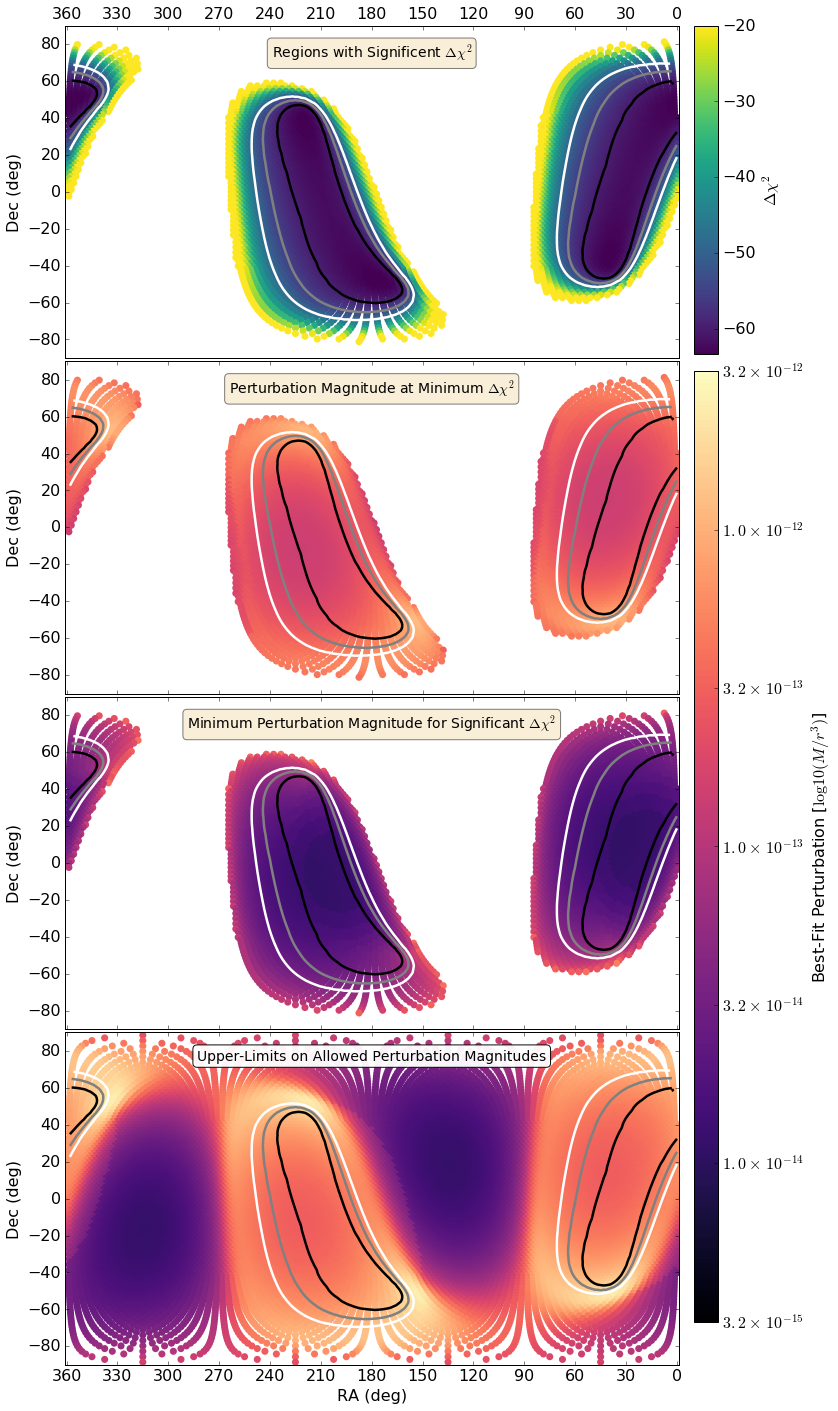}
    \includegraphics[trim = 0mm 370mm 0mm 0mm, clip, angle=0, width=0.49\textwidth]{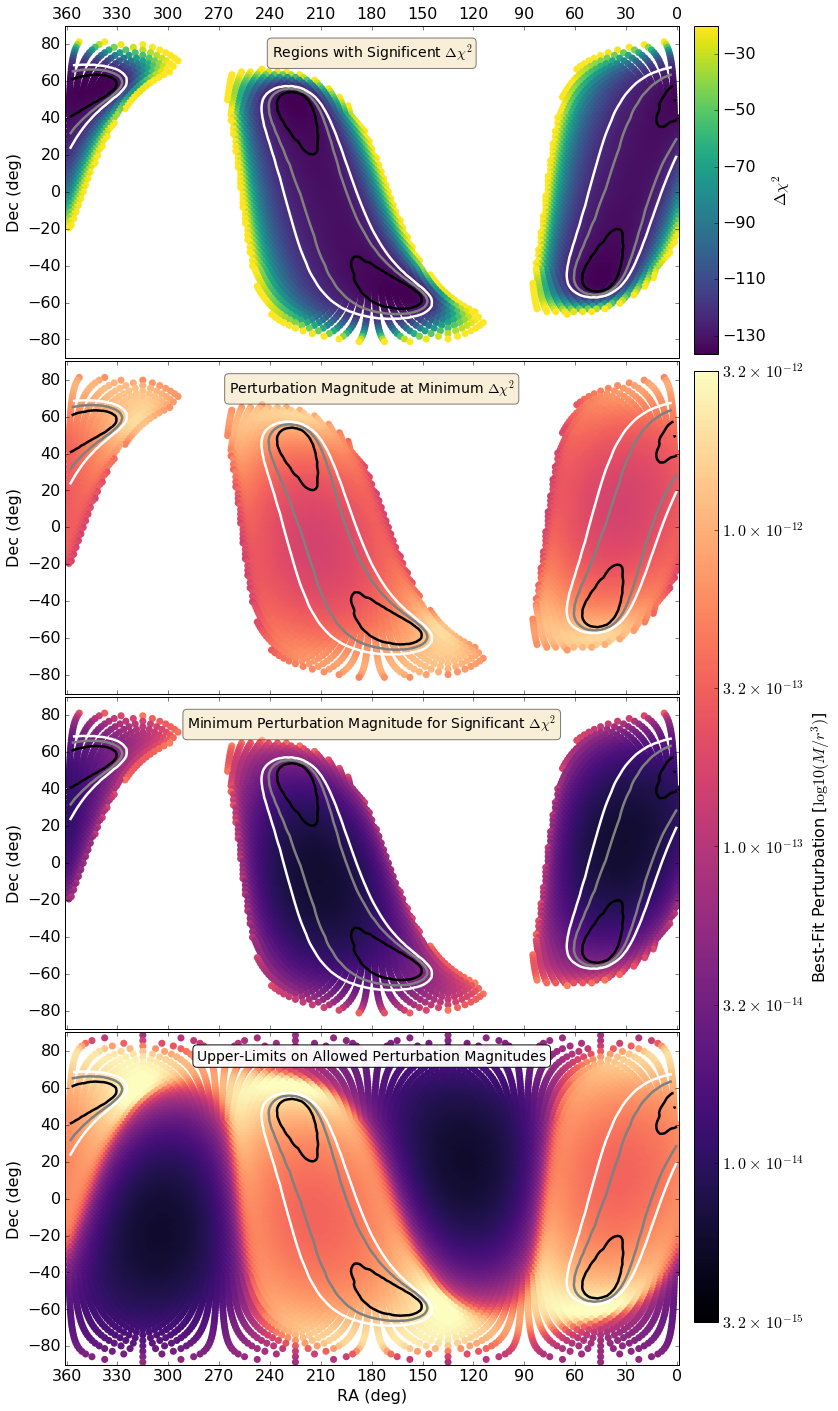}
    \\
    \includegraphics[trim = 0mm 370mm 0mm 0mm, clip, angle=0, width=0.49\textwidth]{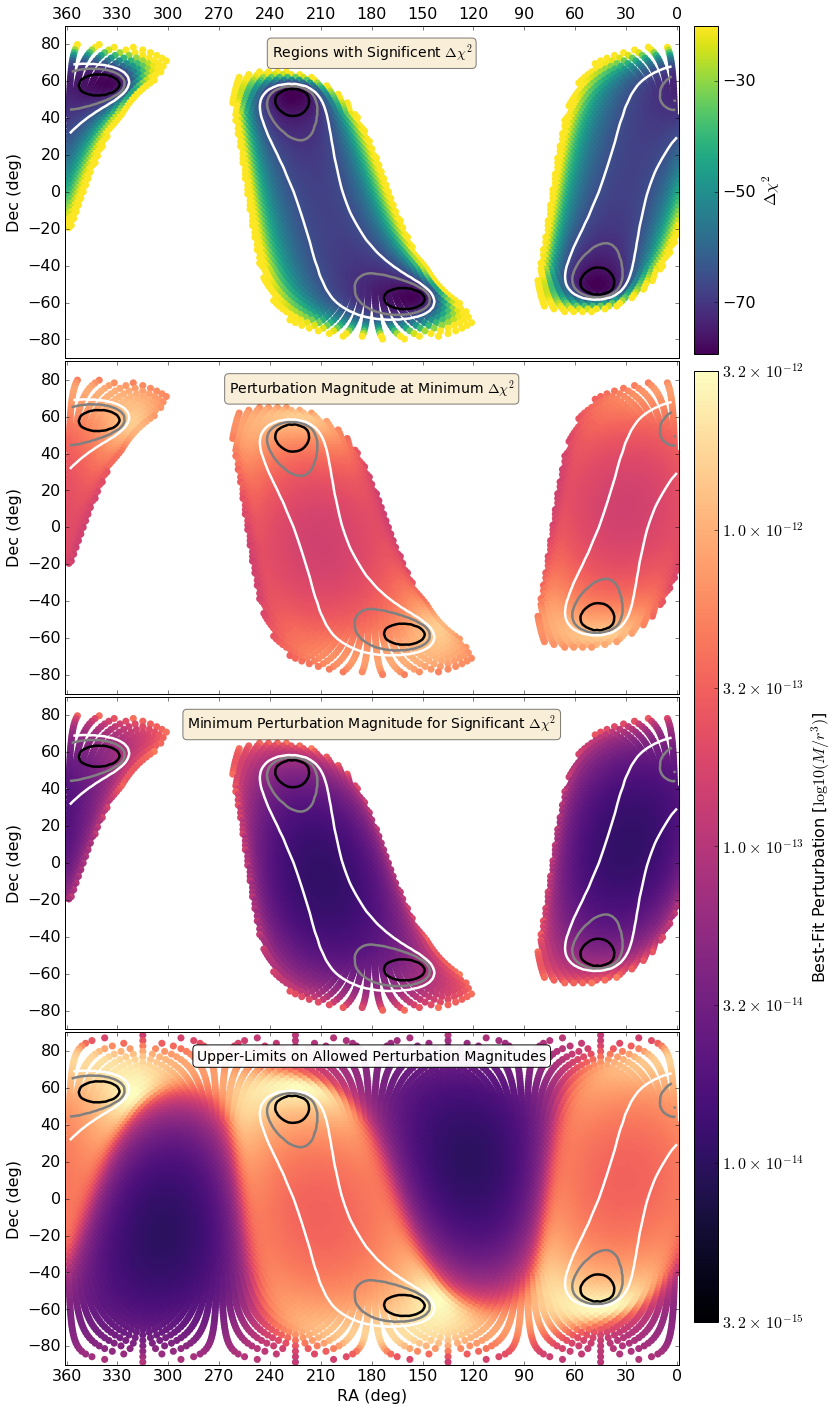}
    \includegraphics[trim = 0mm 370mm 0mm 0mm, clip, angle=0, width=0.49\textwidth]{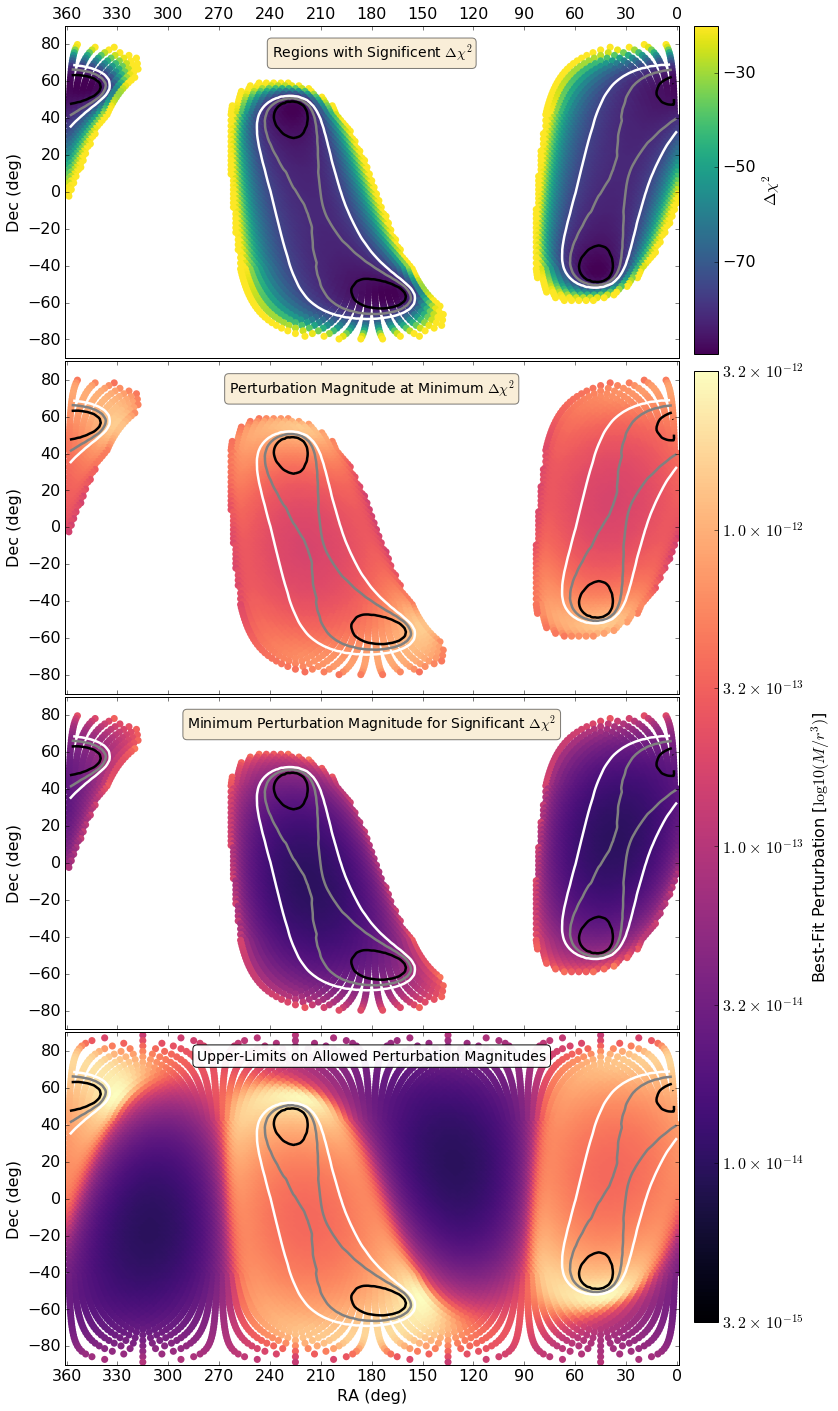}
    \caption{ %
    {\bf Top: } Standard results from Figure \ref{FIG:SKY}
    {\bf Others: } Results from four different block-bootstrap realizations (outliers rejected). 
    The overall results look broadly similar for the different block-bootstrap iterations, with the main difference being in the overall depth / significance of the $\chi^2$ minima (note the different color scales employed for all maps). As seen in the individual bootstrap realizations of Figure \ref{FIG:APP:BOOT}, the locations of the 1 and 2 sigma contours changes somewhat from realization to realization. 
    }
    \label{FIG:APP:BLOCKBOOT}
    \end{minipage}
    \end{figure*}

Figures \ref{FIG:APP:BOOT} and \ref{FIG:APP:BLOCKBOOT} show the respective results of the conventional bootstrap and block-bootstrap analyses described in Section~\ref{SECN:ASSESS}.  For the conventional bootstrap analysis, the overall results look broadly similar for the different bootstrap realizations, with the main difference being in the location of the $1\sigma$ contour.  The results of the block-bootstrap analysis show more variation. 
The overall results look broadly similar for the different block-bootstrap realizations.  However, the overall depth, and thus the significance of the $\chi^2$ minima, varies between realizations (note the different color scales employed for all maps). As seen in the individual bootstrap realizations of Figure \ref{FIG:APP:BOOT}, the locations of the 1 and 2 sigma contours changes somewhat from realization to realization. 

    %

    %


\label{lastpage}
\end{document}